\documentclass[journal]{IEEEtran}

\usepackage{amsmath,amssymb,mathtools,bm,amsthm}
\usepackage{graphicx}
\usepackage{cite}
\usepackage{url}
\usepackage{booktabs}
\usepackage{algorithm}
\usepackage{flafter}
\usepackage{algpseudocode}
\usepackage{textcomp}
\usepackage{enumitem}
\usepackage{multirow}

\theoremstyle{plain}
\newtheorem{theorem}{Theorem}

\newtheorem{proposition}[theorem]{Proposition}
\newtheorem{corollary}[theorem]{Corollary}

\newtheorem{remark}{Remark}
\newtheorem{sty1}{Theorem}
\newtheorem{definition}[sty1]{Definition}

\begin{document}
\title{Embodied Movable Antennas}
\author{Yewen Cao, Dengke Wei, Yulin Shao%
\thanks{The authors are with the Department of Electrical and Computer Engineering, The University of Hong Kong, Hong Kong, China.}
}
\maketitle
\begin{abstract}
This paper introduces embodied movable antennas (MAs), a new MA architecture that rethinks the source of antenna mobility. Unlike conventional MAs that rely on dedicated antenna actuators, embodied MAs harness the mechanical degrees of freedom already present in physical agents to reconfigure antenna position and orientation. We distinguish two modes of embodied MAs: task-coupled and task-preserving. Focusing on the latter, we show how wireless channels can be reconfigured by exploiting task-equivalent host configurations without altering the prescribed physical task. Using a redundant robotic manipulator as a canonical realization, we characterize the intrinsically coupled antenna translation and orientation induced by task-preserving self-motion and derive an exact Fourier representation of the resulting multipath channel. Our analysis reveals that the wireless value of embodied mobility depends not merely on how far an antenna can move, but on which configurations remain feasible under the physical task and how their coupled motion interacts with polarization and multipath propagation. Building on this insight, we develop a cross-task antenna mounting design that enables wireless adaptation with a fixed physical installation. Numerical results demonstrate substantial communication gains through task-preserving motion. These findings establish mechanical redundancy as a new, quantifiable resource for wireless channel reconfiguration.
\end{abstract}


\section{Introduction}
\subsection{Rethinking Antenna Mobility}

Wireless channels are conventionally treated as propagation conditions to be estimated and adapted to. Movable antennas (MAs) \cite{zhu_ma_twc_2024, KKwong_FAS_TWC} introduce a more proactive form of control: by mechanically changing antenna position and orientation, the communication system can reshape the channel itself \cite{shao_6dma_jsac_2025,pama_2025}. Translation alters the relative phases of multipath components \cite{zhu_ma_twc_2024, ding2025flexible}, while orientation changes the directional and polarization response of the antenna~\cite{zheng2025rotatable, pama_2025,shao_6dma_twc_2025}. Together, these mechanisms can promote coherent combining \cite{wu2025globally}, improve polarization matching \cite{pama_2025,wei2026multi}, and create more favorable channel realizations for higher received power, rate, and reliability \cite{zhu_ma_twc_2024,li2026sum}. MA therefore turns antenna pose into a physical control variable of the wireless channel.

\begin{figure}[!t]
    \centering
    \includegraphics[width=1\columnwidth]{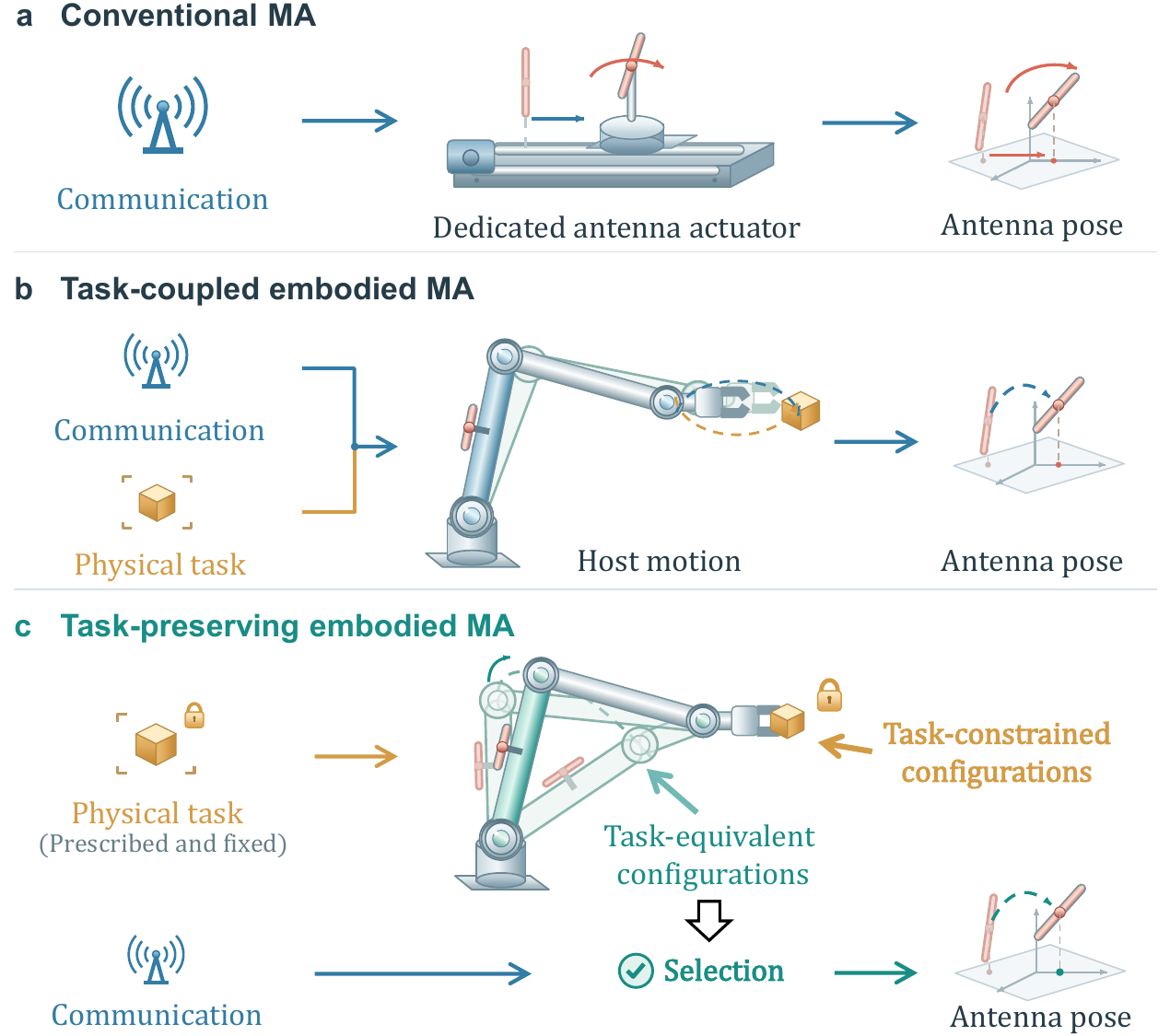}
    \caption{
(a) Conventional MA uses a dedicated antenna actuator to control the antenna pose.
(b) In task-coupled embodied MA, the physical task and communication jointly determine the host motion.
(c) In task-preserving embodied MA, communication selects among task-equivalent host configurations while the prescribed physical task remains unchanged.}
    \label{fig:embodied_ma_architecture}
\end{figure}

Existing MA architectures realize this capability through mechanisms dedicated to antenna reconfiguration~\cite{ning_ma_arch_2025}, such as linear rails~\cite{zhu_ma_tutorial_2025,dong_ma_prototype_2026}, rotary stages~\cite{zheng2025rotatable, zheng_ra_tcom_2026}, or actively deformable structures~\cite{faddoul_2025,Shao2026Embodied,bonilla_eusipco_2019}. This has led to an implicit architectural assumption: \emph{antenna mobility must be separately actuated for communication}. Yet many emerging wireless terminals are themselves mechanically actuated agents \cite{Chen2021WirelessMultirobot,shao2024theory}. A robot, for example, already contains joints and actuators that continuously reposition its body links. If an antenna is integrated with the robot body, those same mechanical degrees of freedom (DoF) can also reposition the antenna. This raises a different possibility: rather than adding mobility to the antenna, can we extract antenna mobility from the host itself?

Motivated by this observation, we propose a new MA architecture, termed the \emph{embodied MA}. An embodied MA is an MA whose position and orientation are reconfigured through the mechanical motion of its host physical agent, without relying on a dedicated antenna actuator. The fundamental change is the source of mobility: a conventional MA introduces mechanical DoF specifically for communication, whereas an embodied MA reuses mechanical DoF already available in the host. Consequently, its feasible antenna poses are no longer determined by antenna hardware alone, but emerge jointly from the host kinematics, physical task, mechanical constraints, and antenna mounting.

We further distinguish two modes of embodied MA based on whether communication is allowed to modify the host motion that accomplishes the physical task, as shown in Fig.~\ref{fig:embodied_ma_architecture}.
\begin{itemize}[leftmargin=0.34cm]
\item In \emph{task-coupled embodied MA}, communication and task jointly determine the host motion. Antenna reconfiguration is therefore obtained by adapting the task motion itself.
\item In \emph{task-preserving embodied MA}, the task is fixed first, and communication exploits only alternative host motions or configurations that leave the prescribed task unchanged.
\end{itemize}

The first mode already appears implicitly in communication-aware robotics \cite{muralidharan_mostofi_2021,bonilla_tutorial_2024}. Robot positions and trajectories have been designed to accomplish physical tasks while accounting for connectivity \cite{zavlanos_proc_ieee_2011,gha_mostofi_tsp_2012}, relay formation \cite{fink_proc_ieee_2012}, or channel quality \cite{gha_mostofi_tac_2011,bonilla_unet_2017}. Viewed through the embodied-MA lens, these approaches use task motion itself as the antenna positioning mechanism. This can expose a broad motion space to wireless optimization, but it also couples communication directly to task planning: a motion favorable to the wireless channel may differ from the motion preferred by the physical task, creating an inherent communication-task tradeoff.

This paper makes the first attempt to systematically exploit the second mode. The key opportunity is \emph{mechanical redundancy}: completing a physical task does not always uniquely determine the configuration of an agent's body~\cite{siciliano_1990,fabregat_2025}. A robot may, for example, keep its hand at exactly the same position and orientation while changing the posture of its arm~\cite{shimizu_tro_2008,elias_wen_2024}. To the task, these postures are equivalent, but to the wireless channel, they are not. If an antenna is mounted on the moving body, such task-equivalent postures can place the antenna at different positions and orientations, thereby creating different channel realizations. Task-preserving embodied MA turns this latent configuration diversity into a wireless reconfiguration resource.

Compared with the task-coupled mode, task-preserving embodied MA deliberately operates within a smaller motion space, but gains a crucial architectural advantage: communication no longer needs to participate in task design. The upper layer remains free to specify the physical task, while the wireless layer adapts only within the resulting set of task-equivalent configurations. This separation enables a modular and task-transparent form of wireless adaptation, allowing the channel to be reshaped without compromising or redesigning the host task.

\subsection{Harnessing Embodied Antenna Mobility}

The promise of task-preserving embodied MA rests on a simple but nontrivial question: whether the residual motion left by a physical task can be turned into predictable and useful wireless reconfiguration. Three properties make this fundamentally different from conventional MA.
First, variations in antenna position and orientation are driven by the same host motion and are therefore intrinsically coupled.
Second, communication can access only the portion of motion that remains feasible under the prescribed task and mechanical constraints, so nominal mobility does not directly translate into usable reconfigurability.
Third, the available antenna motion is itself task dependent: the same physical mounting can generate different feasible antenna pose sets, and hence different wireless capabilities, under different tasks.
Understanding task-preserving embodied MA therefore requires a theory that captures coupling, feasible motion, and task dependence together.

The main contributions of this paper are summarized below.
\begin{itemize}[leftmargin=0.4cm]

\item We establish embodied MA as a new architectural paradigm that expands antenna mobility beyond dedicated positioning hardware to the mechanical DoF already present in physical agents. This shift opens a broader design space in which communication can exploit the host's native mobility at different levels of task interaction. In particular, it makes possible task-transparent wireless adaptation, where the channel can be reconfigured while preserving the autonomy of the upper-layer physical task.

\item We characterize the intrinsic coupling between embodied antenna position and orientation. Starting from a general redundant manipulator, we develop a 7-DoF spherical-revolute-spherical (SRS) manipulator as a canonical realization. With the full end-effector pose fixed, its residual self-motion is parameterized by a swivel angle. We show that the antenna phase center has a circular locus and the dipole axis has a constant cone angle, both parameterized by the same swivel angle. We derive the resulting multipath channel and an exact Fourier representation that retains this coupling, providing a compact bridge from task-preserving body motion to wireless response.

\begin{figure}[!t]
 \centering
 \includegraphics[width=0.7\columnwidth]{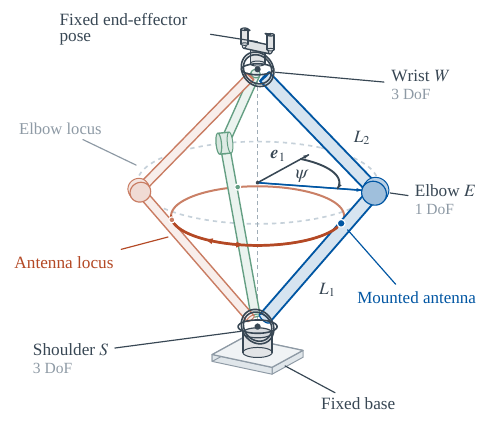}
 \caption{Task-preserving embodied MA with a 7-DoF SRS manipulator. The prescribed end-effector pose is fixed while the swivel motion changes the pose of an antenna fixed to an intermediate link.}
 \label{fig:geometry}
\end{figure}

\item We establish how much wireless reconfigurability remains once the physical task restricts the available motion. Over an arbitrary finite feasible swivel interval, we relate received power variation to the gain attainable through configuration selection. We then resolve the two underlying physical mechanisms analytically. For orientation, we derive a power guarantee valid for arbitrary complex effective fields and obtain in closed form the cone parameter that maximizes this guarantee. For two-path translation, we derive the exact selected power and prove that coherent combining can be guaranteed for every initial phase offset if and only if the attainable relative phase interval spans at least $2\pi$. These results turn residual mechanical freedom into a quantitatively characterized resource for polarization and phase reconfiguration.

\item We address the task dependence of embodied antenna mobility through cross-task mounting design. Because a fixed mounting can induce different feasible antenna pose sets under different prescribed tasks, optimizing the installation for one task need not produce a robust embodiment for another. We therefore optimize one physical mounting over representative tasks and channel statistics, derive a finite-sample formulation with an explicit swivel discretization error bound, and reduce online adaptation to selecting a single feasible configuration for the current task and channel.
\end{itemize}
Numerical results demonstrate that the mechanical freedom remaining after a physical task is specified can provide substantial wireless gains.
At $28$ GHz, with five active paths and a $90^\circ$ feasible swivel span, selecting among task-equivalent robot configurations improves the average received power by $5.30$ dB compared with maintaining a fixed reference configuration.
The results also highlight the importance of accounting for antenna translation and orientation jointly, as optimizing either effect in isolation can lead to substantially inferior configuration choices.
Furthermore, when the robot performs different physical tasks, a single antenna mounting designed across these tasks achieves a 1.64 dB gain over a reference mounting, approaching the 2.02 dB gain attainable by reinstalling the antenna separately for each task. These findings show that the robot's existing mechanical freedom can support effective wireless reconfiguration without altering the prescribed task or requiring dedicated antenna actuation.

\section{System Model}
\label{sec:model}
In task-preserving embodied MA, communication reuses the mechanical freedom of a host instead of moving the antenna with a dedicated actuator. The basic idea is illustrated in Fig.~\ref{fig:geometry}: a robotic arm can keep its hand (end-effector) at the same position and orientation while changing its elbow posture. An antenna attached to the moving arm then changes both its position and orientation, even though the physical task remains unchanged. This section describes which host configurations are available under a fixed task, how a fixed mounting maps them to antenna poses, and which of these poses can be selected for communication.

\subsection{Task-Preserving Self-Motion}
Consider a robotic host with $N_q$ actuated joints and joint coordinates $\bm q=[q_1,\ldots,q_{N_q}]^{\mathsf T}\in\mathbb R^{N_q}$. The joint coordinates determine the poses of the robot links through forward kinematics. A prescribed task $\mathcal T$ constrains the end-effector position, orientation, or both. Let $\mathcal Q_{\mathcal T}\subseteq\mathbb R^{N_q}$ denote one connected set of joint configurations that satisfy the task, joint limits, and collision constraints.

\begin{definition}[Self-motion]
\label{def:task_preserving_self_motion}
For a fixed task $\mathcal T$, a continuous change of the robot configuration within $\mathcal Q_{\mathcal T}$ is called task-preserving self-motion. The internal joint configuration changes, but the prescribed task and the imposed mechanical constraints remain satisfied.
\end{definition}

This freedom arises from kinematic redundancy: a task can constrain the end-effector without fully specifying the robot's internal configuration \cite{siciliano_1990,kreutz_1992}. For task-preserving embodied MA, communication may select only configurations in $\mathcal Q_{\mathcal T}$.

We focus on a fixed-base manipulator performing a static kinematic task that prescribes the full end-effector pose. This specifies three position and three orientation coordinates. The 7-DoF SRS manipulator in Fig.~\ref{fig:geometry} has seven actuated joints: three at the shoulder, one at the elbow, and three at the wrist. At a regular configuration, the six independent end-effector constraints leave one local degree of self-motion. This remaining freedom can be described by the shoulder-elbow-wrist swivel angle $\psi$, which locates the elbow around the shoulder-to-wrist axis~\cite{kreutz_1992,shimizu_tro_2008,elias_wen_2024}. On a selected regular branch, the joint configuration is therefore written as $\bm q=\bm q(\psi)$.

Let $S$, $E$, and $W$ denote the shoulder, elbow, and wrist centers, with positions $\bm p_S$, $\bm p_E$, and $\bm p_W$ expressed in the fixed robot base frame. The upper-arm and forearm lengths are $L_1=\|\bm p_E-\bm p_S\|$ and $L_2=\|\bm p_W-\bm p_E\|$. Since the wrist-to-end-effector transform is fixed, holding the end-effector pose fixed also fixes $\bm p_W$. The elbow must remain at distance $L_1$ from the shoulder and $L_2$ from the wrist. Its possible positions therefore lie at the intersection of two spheres, which is a circle for $|L_1-L_2|<d<L_1+L_2$, where $d=\|\bm p_W-\bm p_S\|$.

To describe this circle, define the unit shoulder-to-wrist axis $\bm e=(\bm p_W-\bm p_S)/d$. Choose a unit reference direction $\bm e_1\perp\bm e$ and $\bm e_2$ such that $(\bm e_1,\bm e_2,\bm e)$ is right-handed. The plane spanned by $(\bm e,\bm e_1)$ sets the zero reference for $\psi$. The elbow position is then
\begin{equation}
 \bm p_E(\psi)=\bm p_S+z_E\bm e
 +r_E\bigl(\bm e_1\cos\psi+\bm e_2\sin\psi\bigr),
 \label{eq:srs_elbow_position}
\end{equation}
where $z_E=(d^2+L_1^2-L_2^2)/(2d)$ is the axial coordinate of the circle center and $r_E=\sqrt{L_1^2-z_E^2}$ is its radius. Equivalently,
\begin{equation}
 r_E^2(d)=\frac{[(L_1+L_2)^2-d^2][d^2-(L_1-L_2)^2]}{4d^2}.
 \label{eq:elbow_radius_task}
\end{equation}
The circle describes the geometrically possible elbow positions, not necessarily the motion that the robot can actually execute. Joint limits and collision constraints restrict the selected regular branch to a feasible swivel interval $\Psi_{\mathcal T}=[\psi_-,\psi_+]$~\cite{shimizu_tro_2008,elias_wen_2024}. We write $\Psi=\Psi_{\mathcal T}$ when the task is fixed. Thus, the communication controller has only one feasible configuration variable, $\psi\in\Psi$.

\begin{figure}[!t]
 \centering
 \includegraphics[width=0.7\columnwidth]{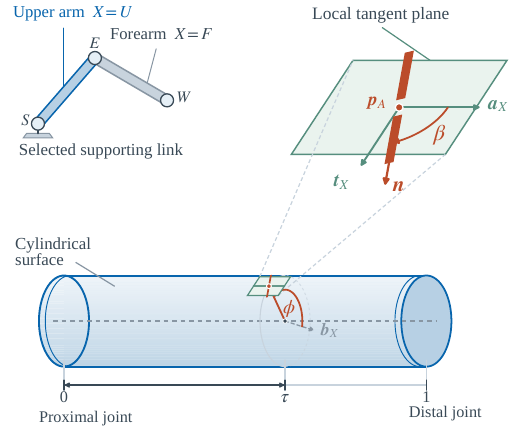}
 \caption{Surface antenna mounting $\bm m=(X,\tau,\phi,\beta)$. The axial coordinate $\tau$ runs from the proximal to distal joint of the selected link. The circumferential angle $\phi$ locates $\bm p_A$ relative to $\bm b_X$. The enlarged tangent plane shows $\beta$, measured from $\bm a_X$ toward $\bm t_X$, and the resulting dipole axis $\bm n$.}
 \label{fig:surface_mounting}
\end{figure}

For later use, define the radial and tangential directions in the elbow circle plane as
\begin{equation*}
 \bm e_r(\psi)=
 \bm e_1\cos\psi+\bm e_2\sin\psi,~
 \bm e_t(\psi)=
 -\bm e_1\sin\psi+\bm e_2\cos\psi.
\end{equation*}
Then~\eqref{eq:srs_elbow_position} becomes $\bm p_E(\psi)=\bm p_S+z_E\bm e+r_E\bm e_r(\psi)$. The same swivel angle determines the poses of the upper arm and forearm, so an antenna mounted on either link inherits this motion.
We assume quasi-static operation: the robot reaches a feasible configuration between communication intervals and holds it during data transmission.

\subsection{Surface Antenna Mounting}
An antenna mounted on the moving upper arm or forearm can change its pose even while the end-effector remains fixed. The antenna motion depends not only on the swivel angle $\psi$ but also on where and how the antenna is attached. As shown in Fig.~\ref{fig:surface_mounting}, we approximate the surface of each supporting link by a circular cylinder of radius $\varrho_X$, where $X\in\{U,F\}$ denotes the upper arm or forearm. This geometry specifies both the attachment point and the local tangent plane used to orient the antenna. The wireless model in Section~\ref{sec:response} uses the phase center and dipole axis of an ideal short dipole. Cylindrical conformal dipoles provide a physical precedent for this mounting model~\cite{fang_qu_tap_2024}.

We describe a physical mounting by $\bm m=(X,\tau,\phi,\beta)$. Here, $X$ chooses the supporting link; $\tau\in[0,1]$ locates the antenna along that link from its proximal to distal joint; $\phi\in[0,2\pi]$ locates it around the cylindrical surface; and $\beta\in[0,\pi]$ sets the dipole direction within the local tangent plane. The angular references are specified below. An ideal short dipole has the same power response for opposite directions along its axis, so $\beta$ has period $\pi$. The mounting variables are fixed after installation: they describe how the antenna is attached, not additional control variables during communication.

Mechanical access, clearance, and cable routing can restrict the admissible mountings, as described in Section~\ref{sec:optimization}. We model the antenna as conformal with negligible thickness in collision checking and assume the admissible mounting does not further restrict the robot self-motion. Under this assumption, $\Psi$ depends on the task and robot constraints but not on $\bm m$.

To express the antenna pose in the fixed robot base frame, we first define three perpendicular directions attached to the selected link: one along the link axis and two transverse to it. Let $\sigma_U=1$ and $\sigma_F=-1$ distinguish the two links, and define
\begin{equation*}
 c_U\triangleq\frac{z_E}{L_1},~~
 s_U\triangleq\frac{r_E}{L_1},~~
 c_F\triangleq\frac{d-z_E}{L_2},~~
 s_F\triangleq\frac{r_E}{L_2}.
\end{equation*}
These coefficients satisfy $c_X^2+s_X^2=1$. The unit link-axis direction $\bm a_X(\psi)$ and a transverse direction $\bm b_X(\psi)$ in the arm plane are
\begin{equation*}
 \bm a_X(\psi)=c_X\bm e+\sigma_Xs_X\bm e_r(\psi),~~
 \bm b_X(\psi)=c_X\bm e_r(\psi)-\sigma_Xs_X\bm e.
\end{equation*}
Together with $\bm e_t(\psi)$, they form a local orthonormal frame that follows the link as the elbow swivels. The centerline point selected by $\tau$, expressed in the fixed base frame, is
\begin{equation*}
\begin{aligned}
 \bar{\bm p}_X(\psi)
 &=\bm p_S+z_{0,X}\bm e+\rho_{0,X}\bm e_r(\psi),\\
 (z_{0,U},\rho_{0,U})&=(\tau z_E,\tau r_E),\\
 (z_{0,F},\rho_{0,F})&=
 \bigl(z_E+\tau(d-z_E),(1-\tau)r_E\bigr).
\end{aligned}
\end{equation*}

The circumferential angle $\phi$ is measured from $\bm b_X(\psi)$ toward $\bm e_t(\psi)$. At this angle, the outward surface normal and circumferential tangent are
\begin{equation*}
\begin{aligned}
 \bm r_X(\psi,\phi)
 &=\cos\phi\,\bm b_X(\psi)+\sin\phi\,\bm e_t(\psi),\\
 \bm t_X(\psi,\phi)
 &=-\sin\phi\,\bm b_X(\psi)+\cos\phi\,\bm e_t(\psi).
\end{aligned}
\end{equation*}
The dipole angle $\beta$ is measured in the local tangent plane from the link axis $\bm a_X(\psi)$ toward $\bm t_X(\psi,\phi)$.

For the ideal short dipole considered in this paper, the phase center is taken at the dipole center and placed at the selected point on the cylindrical surface. Its orientation is fully characterized by the unit dipole axis $\bm n\in\mathbb R^3$, with $\lVert\bm n\rVert=1$. Rotation about this axis does not change the response of an ideal short dipole. The resulting phase center and dipole axis are
\begin{equation}
\begin{aligned}
 \bm p_A(\psi;\bm m)
 &=\bar{\bm p}_X(\psi)+\varrho_X\bm r_X(\psi,\phi),\\
 \bm n(\psi;\bm m)
 &=\cos\beta\,\bm a_X(\psi)+\sin\beta\,\bm t_X(\psi,\phi).
\end{aligned}
 \label{eq:surface_mounting_pose}
\end{equation}
Thus, $(X,\tau,\phi)$ specify the attachment point on the link, and $\beta$ specifies the dipole direction relative to its tangent plane.

\begin{remark}[Fixed mounting versus online reconfiguration]
\label{rem:fixed_mounting}
The mounting $\bm m$ is a physical design choice made before operation. For a prescribed task, mounting design determines the antenna pose trajectory, while online adaptation selects a configuration along that trajectory through $\psi\in\Psi$. Section~\ref{sec:optimization} addresses how to choose one mounting across tasks.
\end{remark}

\subsection{Coupled Antenna Motion}
Equation~\eqref{eq:surface_mounting_pose} determines the antenna pose for each feasible robot configuration. Unlike a conventional MA with independently actuated translation and rotation, the antenna position and orientation here are both controlled by the same swivel angle $\psi$. Consequently, a position reached at one swivel angle cannot, in general, be combined with an orientation reached at another. The following result provides a compact geometric description of this coupled motion.

\begin{proposition}[SRS antenna motion]
\label{prop:srs_pose}
For any fixed mounting $\bm m$ and regular SRS branch, the phase center and dipole axis can be expressed in the basis fixed by the prescribed task $(\bm e_1,\bm e_2,\bm e)$ as
\begin{equation}
\begin{aligned}
 \bm p_A(\psi;\bm m)=\bm p_c+\rho\bigl[
 \bm e_1\cos(\psi+\delta_p) +\bm e_2\sin(\psi+\delta_p)\bigr],
\end{aligned}
 \label{eq:srs_position_locus}
\end{equation}
\begin{equation}
\begin{aligned}
 \bm n(\psi;\bm m)=\nu\bm e+\sqrt{1-\nu^2}\bigl[
 \bm e_1\cos(\psi+\delta_n)\\
 \qquad+\bm e_2\sin(\psi+\delta_n)\bigr].
\end{aligned}
 \label{eq:srs_orientation_motion}
\end{equation}
The phase center is specified by
\begin{equation}
\begin{aligned}
 \rho&=
 \sqrt{\bigl(\rho_{0,X}+\varrho_Xc_X\cos\phi\bigr)^2
 +\varrho_X^2\sin^2\phi},\\
 \delta_p&=\operatorname{atan2}
 \bigl(\varrho_X\sin\phi,\rho_{0,X}+\varrho_Xc_X\cos\phi\bigr),\\
 \bm p_c&=\bm p_S+
 \bigl(z_{0,X}-\sigma_X\varrho_Xs_X\cos\phi\bigr)\bm e,
\end{aligned}
 \label{eq:surface_position_descriptors}
\end{equation}
The dipole orientation is specified by
\begin{equation}
\begin{aligned}
 \nu&= c_X\cos\beta
 +\sigma_Xs_X\sin\phi\sin\beta,\\
 \delta_n&=\operatorname{atan2}\!\left(
 \cos\phi\sin\beta,
 \sigma_Xs_X\cos\beta-c_X\sin\phi\sin\beta
 \right).
\end{aligned}
 \label{eq:surface_orientation_descriptors}
\end{equation}
If $\rho=0$, set $\delta_p=0$. If $|\nu|=1$, set $\delta_n=0$. The corresponding angular offset is immaterial when the associated transverse component vanishes.
\end{proposition}
\begin{proof}
    See the supplementary material.
\end{proof}

Proposition~\ref{prop:srs_pose} is the interface between the physical mounting and the wireless model. Equation~\eqref{eq:srs_position_locus} parameterizes a circle centered at $\bm p_c$ in the plane normal to $\bm e$, with radius $\rho$, while $\delta_p$ specifies the angular offset relative to $\psi$. Restricting $\psi$ to $\Psi$ gives the feasible phase center locus, which is the corresponding circular arc. The dipole axis has constant axial component $\nu$ and rotates about $\bm e$ with angular offset $\delta_n$, and the same feasible interval restricts its accessible orientation range.

The task and mounting jointly determine $\bm p_c$, $\rho$, $\delta_p$, $\nu$, and $\delta_n$, while joint limits and collision constraints restrict the feasible swivel interval $\Psi$. Thus, the antenna configurations available for communication can differ across tasks even when the physical mounting $\bm m$ remains unchanged.

\section{Channel Response under Self-Motion}
\label{sec:response}

Section~\ref{sec:model} established that, for a prescribed task and a fixed antenna mounting, the antenna position and orientation are both determined by a single feasible swivel angle $\psi\in\Psi$. We now examine how this task-preserving motion changes the wireless channel. In particular, antenna translation modifies the propagation phases of multipath components, while antenna orientation changes their polarization responses.
This section develops the resulting channel model and its harmonic representation, which will be used to quantify wireless reconfigurability in Section~\ref{sec:reconfigurability}.

\subsection{Multipath Channel Model}

Consider a narrowband far-field channel with $L$ propagation paths \cite{zhu_ma_twc_2024,zhu_ma_tutorial_2025}. All spatial vectors are expressed in the fixed robot base frame. At carrier wavelength $\lambda_c$, let $k=2\pi/\lambda_c$ denote the wavenumber.

Each path $\ell$ is characterized by a unit propagation direction $\bm u_\ell\in\mathbb R^3$, a unit polarization vector $\bm\epsilon_\ell\in\mathbb C^3$, and a complex path coefficient $\bar\alpha_\ell$ referenced to the robot base origin. The propagation direction is defined such that an antenna displacement $\Delta\bm p$ introduces the phase factor $e^{-jk\bm u_\ell^{\mathsf T}\Delta\bm p}$. The multipath channel realization is described by $\mathcal H=\{\bar\alpha_\ell,\bm u_\ell,\bm\epsilon_\ell\}_{\ell=1}^{L}$, with $\|\bm\epsilon_\ell\|=1$ and $\bm u_\ell^{\mathsf T}\bm\epsilon_\ell=0$.

As in conventional local MA channel models, we adopt a plane-wave approximation over the feasible antenna motion region~\cite{zhu_ma_twc_2024,zhu_ma_tutorial_2025}. The propagation parameters, i.e., $\bm u_\ell$, $\bm\epsilon_\ell$, and $\bar\alpha_\ell$, are assumed unchanged as the antenna moves. We further assume that the propagation environment remains quasi-static during swivel selection and the subsequent communication interval.

To model the effect of antenna orientation on reception, we consider an ideal short electric dipole whose receive response is proportional to the projection of the incident electric field onto its axis~\cite{balanis_2016,pama_2025}. This model neglects finite-length effects, electromagnetic coupling with the robot body, and variations in antenna loading. After absorbing the fixed effective length into $\bar\alpha_\ell$, the response to path $\ell$ is $g_\ell(\psi;\bm m)
    =\bm n(\psi;\bm m)^{\mathsf T}\bm\epsilon_\ell$,
where $\bm n(\psi;\bm m)$ is the dipole axis given in~\eqref{eq:srs_orientation_motion}.

The resulting complex channel coefficient is
\begin{equation}
    h(\psi;\bm m,\mathcal H)
    =\sum_{\ell=1}^{L}\bar\alpha_\ell
    g_\ell(\psi;\bm m)
    e^{-jk\bm u_\ell^{\mathsf T}\bm p_A(\psi;\bm m)}.
    \label{eq:mounting_channel_definition}
\end{equation}
With unit transmit power, the received power is
$P(\psi;\bm m,\mathcal H)=|h(\psi;\bm m,\mathcal H)|^2$.
For fixed transmit power and noise variance, maximizing $P$ is equivalent to maximizing the receive signal-to-noise ratio (SNR).

\subsection{Coupled Position-Orientation Response}

We next make the dependence on the swivel angle explicit. For fixed $(\bm m,\mathcal H)$, we omit these arguments from $h$, $g_\ell$, and $P$ for notational simplicity.

First, the conical orientation motion in~\eqref{eq:srs_orientation_motion} determines the polarization response $g_\ell(\psi)$. The axial component $\nu$ and transverse component $\sqrt{1-\nu^2}$ determine how the dipole axis projects onto the incident field, while $\delta_n$ specifies the angular offset of its transverse orientation.

Next, we consider the position-dependent propagation phase by substituting the circular trajectory in~\eqref{eq:srs_position_locus}. Define the path coefficient referenced to the circle center $\bm p_c$ as
$\alpha_\ell=\bar\alpha_\ell e^{-jk\bm u_\ell^{\mathsf T}\bm p_c}$,
and introduce
\begin{equation*}
\begin{aligned}
    \kappa_\ell
    &\triangleq k\rho\sqrt{
    (\bm u_\ell^{\mathsf T}\bm e_1)^2+
    (\bm u_\ell^{\mathsf T}\bm e_2)^2},\\
    \zeta_\ell
    &\triangleq \operatorname{atan2}
    (\bm u_\ell^{\mathsf T}\bm e_2,
     \bm u_\ell^{\mathsf T}\bm e_1).
\end{aligned}
\end{equation*}
Set $\zeta_\ell=0$ when $\kappa_\ell=0$. Here, $\kappa_\ell$ quantifies the amplitude of the swivel-induced propagation phase variation, in radians, and $\zeta_\ell$ describes the projected path direction in the swivel plane. In particular, $\kappa_\ell=0$ when the path direction is parallel to the shoulder-to-wrist axis, since the circular antenna motion has no displacement component along that direction.

With these definitions, the channel in~\eqref{eq:mounting_channel_definition} becomes
\begin{equation}
    h(\psi)=\sum_{\ell=1}^{L}\alpha_\ell
    g_\ell(\psi)
    e^{-j\kappa_\ell\cos(\psi+\delta_p-\zeta_\ell)}.
    \label{eq:coupled_channel_swivel}
\end{equation}
Expanding the received power separates the individual-path powers from the cross-path terms:
\begin{equation}
\begin{aligned}
    P(\psi)
    ={}&\sum_{\ell=1}^{L}
    |\alpha_\ell|^2|g_\ell(\psi)|^2\\
    &+2\operatorname{Re}
    \sum_{1\le\ell<i\le L}
    \alpha_\ell\alpha_i^*
    g_\ell(\psi)g_i^*(\psi)
    e^{-j\vartheta_{\ell i}(\psi)},
\end{aligned}
\label{eq:power_decomposition}
\end{equation}
where $\vartheta_{\ell i}(\psi)
    =k(\bm u_\ell-\bm u_i)^{\mathsf T}
    [\bm p_A(\psi)-\bm p_c]$
is the relative propagation phase induced by antenna translation between paths $\ell$ and $i$.

\begin{remark}[Polarization and multipath phase]
\label{rem:mechanism_roles}
Equation~\eqref{eq:power_decomposition} shows that antenna orientation affects both the individual-path powers and the cross-path terms through $g_\ell(\psi)$, so it can change received power even in a single-path channel.
Antenna translation affects received power only through the relative propagation phases between paths.
Position-dependent phase control has been studied in conventional MA~\cite{zhu_ma_twc_2024}, while orientation-dependent polarization matching has been investigated in polarization-aware MA~\cite{pama_2025}.
Here, both effects are governed by the same $\psi$, and a configuration favorable for polarization matching need not favor multipath phase combining.
\end{remark}

\subsection{Harmonic Structure of the Channel}

The coupled channel in~\eqref{eq:coupled_channel_swivel} has a particular mathematical structure inherited from the circular position motion and conical orientation motion. This structure is useful for characterizing how much the received power can vary when only a limited swivel interval is available.

To expose it, first consider the orientation response. From~\eqref{eq:srs_orientation_motion}, $g_\ell(\psi)$ is a sinusoidal function of $\psi$ and contains only the Fourier orders $-1$, $0$, and $1$: $g_\ell(\psi)=
    \sum_{s=-1}^{1}o_{\ell,s}e^{js\psi}$,
where
\begin{equation*}
\begin{aligned}
    o_{\ell,0}
    &=\nu\bm e^{\mathsf T}\bm\epsilon_\ell,\\
    o_{\ell,+1}
    &=\frac{\sqrt{1-\nu^2}}{2}
    \bigl(\bm e_1^{\mathsf T}\bm\epsilon_\ell
    -j\bm e_2^{\mathsf T}\bm\epsilon_\ell\bigr)
    e^{j\delta_n},\\
    o_{\ell,-1}
    &=\frac{\sqrt{1-\nu^2}}{2}
    \bigl(\bm e_1^{\mathsf T}\bm\epsilon_\ell
    +j\bm e_2^{\mathsf T}\bm\epsilon_\ell\bigr)
    e^{-j\delta_n}.
\end{aligned}
\end{equation*}

Translation has a different harmonic structure. Although the phase center follows a circle, its contribution to the channel appears inside a complex exponential in~\eqref{eq:coupled_channel_swivel}. Applying the Jacobi-Anger expansion \cite{olver2010nist} to this exponential and combining it with the orientation response yields an exact Fourier representation of the complete channel.

\begin{theorem}[Channel and Power Fourier Representation]
\label{thm:coupled_fourier_representation}
The channel in~\eqref{eq:coupled_channel_swivel} admits the exact, absolutely convergent Fourier representation
$h(\psi)=\sum_{n=-\infty}^{\infty}H_ne^{jn\psi}$,
where $J_r(\cdot)$ is the Bessel function of the first kind of integer order $r$, and
\begin{equation}
    H_n=\sum_{\ell=1}^{L}\alpha_\ell
    \sum_{s=-1}^{1}o_{\ell,s}
    (-j)^{n-s}J_{n-s}(\kappa_\ell)
    e^{j(n-s)(\delta_p-\zeta_\ell)}.
    \label{eq:channel_fourier_coefficients}
\end{equation}
The received power consequently admits the exact, absolutely convergent representation
\begin{equation}
\begin{aligned}
    P(\psi)&=\sum_{m=-\infty}^{\infty}
    C_me^{jm\psi},
    \qquad C_{-m}=C_m^*,\\
    C_m&=\sum_{n=-\infty}^{\infty}
    H_{n+m}H_n^*,
    \qquad m\in\mathbb Z.
\end{aligned}
\label{eq:power_fourier_series}
\end{equation}
If $\rho=0$, then $H_n=0$ for $|n|>1$ and hence $C_m=0$ for $|m|>2$.
\end{theorem}
\begin{proof}
    See the supplementary material.
\end{proof}

Theorem~\ref{thm:coupled_fourier_representation} provides a compact description of how embodied antenna motion shapes the wireless channel. In~\eqref{eq:channel_fourier_coefficients}, the three orientation coefficients $o_{\ell,s}$ are convolved with the Bessel coefficients associated with circular translation. Orientation alone produces at most three channel harmonics, whereas translation can generate higher-order harmonics through $J_{n-s}(\kappa_\ell)$. Equation~\eqref{eq:power_fourier_series} then expresses the received-power harmonics through the autocorrelation of the channel coefficients.

Importantly, the Fourier coefficients retain the physical parameters of the resulting antenna motion. The circle center $\bm p_c$ determines the reference phases through $\alpha_\ell$, while the radius $\rho$ controls the translation-induced phase variations through $\kappa_\ell$. The signed axial component $\nu$ determines the balance between axial and transverse dipole responses. The offsets $\delta_p$ and $\delta_n$ specify how the position and orientation trajectories are aligned with each other and with the propagation directions. Thus, the same mounting can produce different channel responses when the prescribed task changes these motion parameters.

\begin{remark}[Harmonic representation versus feasible motion]
The Fourier representation describes the $2\pi$-periodic channel function generated by the circular and conical motion equations.
For a prescribed task, the robot's mechanical constraints restrict the accessible swivel angle to $\psi\in\Psi$, selecting the corresponding portion of this periodic response.
Consequently, the wireless reconfigurability is jointly determined by the harmonic coefficients $C_m$ and the feasible interval $\Psi$.
This motivates the finite-interval analysis in Section~\ref{sec:reconfigurability}.
\end{remark}

\section{Wireless Reconfigurability under Task Constraints}
\label{sec:reconfigurability}
The channel model in Section~\ref{sec:response} describes how task-preserving self-motion changes the wireless response. However, the existence of antenna motion does not necessarily imply a useful communication gain. A prescribed task may leave only a limited range of robot configurations, and the channel may vary little over that range. We therefore examine how much wireless reconfiguration can be achieved by selecting among the configurations that remain feasible.

Consider a prescribed task, a fixed mounting $\bm m$, and channel $\mathcal H$. The communication controller can select the swivel angle only within $\Psi=[\psi_-,\psi_+]$. We consider a feasible interval spanning at most one swivel revolution and define $\psi_c\triangleq(\psi_-+\psi_+)/2$ and $\Delta\triangleq(\psi_+-\psi_-)/2\in(0,\pi]$, where $\psi_c$ is the interval center and $2\Delta$ is the available swivel span. The centered coordinate $\tilde{\psi}=\psi-\psi_c$ satisfies $\tilde{\psi}\in[-\Delta,\Delta]$. With $\bm m$ and $\mathcal H$ fixed, we write $P(\psi)$ for $P(\psi;\bm m,\mathcal H)$ throughout this section.

Two aspects of the embodied antenna motion determine its reconfiguration capability: how the antenna pose changes with $\psi$, and how much of that motion the prescribed task permits. To quantify the first aspect, recall the position radius $\rho$ and the signed axial orientation component $\nu$ in Proposition~\ref{prop:srs_pose}. Define $\mu\triangleq\nu^2\in[0,1]$.
The quantity $\sqrt{\mu}=|\bm e^{\mathsf T}\bm n|$ measures the axial component of the dipole direction, while $\sqrt{1-\mu}$ is the transverse radius of its orientation cone. Equations~\eqref{eq:srs_position_locus} and~\eqref{eq:srs_orientation_motion} give $\|d\bm p_A/d\psi\|=\rho$ and $\|d\bm n/d\psi\|=\sqrt{1-\mu}$.
Thus, $\rho$ and $\mu$ characterize the magnitudes of antenna translation and orientation change per unit swivel. The resulting wireless gain further depends on the feasible interval $\Psi$ and how the antenna motion interacts with the propagation environment.

We first characterize the selectable received power using the complete coupled channel. We then examine orientation and translation separately to identify their respective wireless capabilities.

\subsection{Selectable Power over Feasible Self-Motion}

For a fixed task and channel realization, task-preserving self-motion provides multiple feasible antenna poses and hence multiple received-power values. We quantify the communication benefit of pose selection by comparing the maximum received power with its average over the feasible interval.

\begin{definition}[Received-power measures over a feasible interval]
\label{def:feasible_power_variation}
For a feasible swivel interval $\Psi$, the maximum received power available through configuration selection is $P^\star\triangleq\max_{\psi\in\Psi}P(\psi)$.
The mean and variance of the received power over the same interval are $\bar P_\Psi \triangleq\frac{1}{2\Delta}\int_\Psi P(\psi)\,d\psi$
    and
    $\sigma_{P,\Psi}^2\triangleq\frac{1}{2\Delta}\int_\Psi
    [P(\psi)-\bar P_\Psi]^2\,d\psi$.
\end{definition}

The mean $\bar P_\Psi$ provides a reference for evaluating configuration selection, while $\sigma_{P,\Psi}^2$ measures the channel variation exposed by the feasible self-motion. If the received power is constant over $\Psi$, selecting a different posture cannot improve it. Otherwise, some feasible configurations provide higher received power than others.

To bound the available improvement, we first characterize how rapidly the received power can change with the swivel angle. For each pair of propagation paths, define
    $r_{\ell i}
    \triangleq k\rho\left\|
    (\bm I-\bm e\bm e^{\mathsf T})
    (\bm u_\ell-\bm u_i)
    \right\|$,
where $\bm I$ is the $3\times3$ identity matrix. The quantity $r_{\ell i}$ is the amplitude, in radians, of the relative propagation-phase variation induced by circular antenna translation. It depends on the antenna motion radius and the difference between the two propagation directions projected onto the swivel plane.

Using the coupled power response in~\eqref{eq:power_decomposition}, an upper bound on its slope is
\begin{equation}
 B_P=
 2\sqrt{1-\mu}\left(\sum_{\ell=1}^{L}|\alpha_\ell|\right)^2
 +2\sum_{1\le\ell<i\le L}|\alpha_\ell\alpha_i|r_{\ell i}.
 \label{eq:power_slope_bound}
\end{equation}
Specifically, $|dP(\psi)/d\psi|\le B_P$ for all $\psi$. The first term bounds the power variation caused by antenna orientation, while the second bounds the variation caused by multipath phase changes. The bound therefore accounts for both mechanisms even though they are driven by the same swivel angle.

\begin{proposition}[Power variation over a feasible swivel interval]
\label{prop:power_reconfigurability}
Let $C_m$ be the power Fourier coefficients in~\eqref{eq:power_fourier_series}, and let $\operatorname{sinc}(x)\triangleq\sin(x)/x$, with $\operatorname{sinc}(0)=1$. Then
\begin{equation*}
    \bar P_\Psi
    =\sum_{m=-\infty}^{\infty}
    C_m e^{jm\psi_c}
    \operatorname{sinc}(m\Delta).
\end{equation*}
For every feasible interval of positive length, $\sigma_{P,\Psi}^2=0$ if and only if $C_m=0$ for all $m\neq0$.
If $\bar P_\Psi>0$, the improvement attainable through configuration selection satisfies
\begin{equation}
    \frac{\sigma_{P,\Psi}^2}{\bar P_\Psi}
    \le P^\star-\bar P_\Psi
    \le\Delta B_P.
    \label{eq:selection_gain_bounds}
\end{equation}
For a full $2\pi$ swivel range,
$\bar P_{2\pi}=C_0$ and
$\sigma_{P,2\pi}^2=\sum_{m\neq0}|C_m|^2$.
\end{proposition}
\begin{proof}
    See the supplementary material.
\end{proof}

Proposition~\ref{prop:power_reconfigurability} connects channel variation over the feasible configurations to the received power attainable through selection. The lower bound shows that a nonzero power variance guarantees an improvement over the interval mean. The upper bound shows that this improvement cannot exceed $\Delta B_P$. In particular, even when the antenna motion can produce a substantial channel variation over a complete swivel revolution, a small feasible interval may expose only a limited part of that variation.

\begin{remark}[Configuration Variation versus Selection Gain]
\label{rem:selection_opportunity}
The received-power variation over feasible configurations determines the opportunity for pose selection, while the resulting gain depends on the reference configuration. In particular, $P^\star-\bar P_\Psi$ measures the improvement relative to the interval average, whereas the improvement relative to a fixed feasible pose $\psi_0$ is $P^\star-P(\psi_0)$. The latter can vanish even when $\sigma_{P,\Psi}^2>0$.
\end{remark}

The preceding result applies to the complete channel, where translation and orientation are coupled. To understand what each mechanism contributes, we next consider two analytical cases that isolate their effects.

\subsection{Polarization Reconfiguration through Orientation}

We first examine the wireless capability provided by antenna orientation. To isolate this mechanism, suppress the translational phase variation by setting $\rho=0$ while retaining the conical dipole motion in Proposition~\ref{prop:srs_pose}. The multipath components then combine into a single complex effective electric-field vector $\bm f=\sum_{\ell=1}^{L}
    \alpha_\ell\bm\epsilon_\ell$.
The received power becomes $P(\tilde{\psi})
    =|\bm n(\tilde{\psi})^{\mathsf T}\bm f|^2$.
Here, $\|\bm f\|^2$ represents the power scale of the effective field before projection onto the receiving dipole. The received power depends on how well the dipole direction aligns with this field.

In a conventional polarization-aware MA, antenna orientation can be adjusted to improve polarization matching~\cite{pama_2025}. In task-preserving embodied MA, however, the antenna cannot take an arbitrary orientation: its dipole axis is restricted to the cone and feasible swivel interval determined by the task and mounting. We therefore ask how much received power can be guaranteed despite this restricted orientation range.

Using the sign symmetry of an ideal short dipole, choose the dipole-axis representative with a nonnegative axial component. We can rotate the transverse basis so that the dipole has zero azimuth at $\tilde{\psi}=0$. In this coordinate system,
\begin{equation*}
    \bm n(\tilde{\psi})
    =
    \begin{bmatrix}
        \sqrt{1-\mu}\cos\tilde{\psi}\\
        \sqrt{1-\mu}\sin\tilde{\psi}\\
        \sqrt{\mu}
    \end{bmatrix},
    \qquad |\tilde{\psi}|\le\Delta,
\end{equation*}
where the third basis direction is $\bm e$.

To describe the orientation coverage over the feasible interval, define the second-moment matrix
\begin{equation*}
    \bm Q_\Delta(\mu)
    \triangleq\frac{1}{2\Delta}
    \int_{-\Delta}^{\Delta}
    \bm n(\tilde{\psi})
    \bm n(\tilde{\psi})^{\mathsf T}
    d\tilde{\psi}.
\end{equation*}
This matrix summarizes how the feasible dipole directions project onto different electric-field directions. In particular, its smallest eigenvalue characterizes the weakest average polarization response over all possible effective-field directions.

Let $S_\Delta=\operatorname{sinc}(\Delta)$, $U_\Delta=[1+\operatorname{sinc}(2\Delta)]/2$, and $V_\Delta=[1-\operatorname{sinc}(2\Delta)]/2$. Direct integration gives
\begin{equation*}
 \bm Q_\Delta(\mu)=
 \begin{bmatrix}
 (1-\mu)U_\Delta & 0 & \sqrt{\mu(1-\mu)}S_\Delta\\
 0 & (1-\mu)V_\Delta & 0\\
 \sqrt{\mu(1-\mu)}S_\Delta & 0 & \mu
 \end{bmatrix}.
\end{equation*}

\begin{theorem}[Orientation guarantee over a finite interval]
\label{thm:orientation_coverage}
For every nonzero complex effective field $\bm f$, the maximum received power attainable through feasible dipole rotation satisfies
\begin{equation}
    \frac{P^\star}{\|\bm f\|^2}
    \ge\lambda_\Delta(\mu),
    \label{eq:orientation_coverage_bound}
\end{equation}
where $\lambda_\Delta(\mu)=\lambda_{\min}(\bm Q_\Delta(\mu))$ and
\begin{equation}
\begin{aligned}
    \lambda_\Delta(\mu)
    &=\min\bigl\{(1-\mu)V_\Delta,
        \lambda_-(\mu,\Delta)\bigr\},\\
    \lambda_-(\mu,\Delta)
    &=\frac{1}{2}\Bigl(
    (1-\mu)U_\Delta+\mu\\
    &\quad-\sqrt{
    [(1-\mu)U_\Delta-\mu]^2
    +4\mu(1-\mu)S_\Delta^2}
    \Bigr).
\end{aligned}
\label{eq:orientation_coverage_eigenvalue}
\end{equation}
For $0<\mu<1$ and $\Delta>0$,
$\lambda_\Delta(\mu)>0$.

Over a full $2\pi$ swivel range, the bound is the exact worst-case value over all complex effective fields:
\begin{equation}
    \min_{\bm f\neq\bm0}
    \frac{P^\star}{\|\bm f\|^2}
    =\lambda_\pi(\mu)
    =\min\left\{\mu,\frac{1-\mu}{2}\right\}.
    \label{eq:full_circle_complex_capability}
\end{equation}
\end{theorem}
\begin{proof}
    See the supplementary material.
\end{proof}

Theorem~\ref{thm:orientation_coverage} establishes a received-power guarantee that holds without specifying the effective-field polarization. The guarantee depends on the cone parameter $\mu$ and the feasible swivel span $2\Delta$. A small $\mu$ provides substantial transverse orientation variation but weak axial coverage, whereas a large $\mu$ strengthens the axial component at the expense of transverse variation. Neither extreme offers a strong guarantee against arbitrary effective-field directions. This tradeoff suggests choosing an intermediate cone parameter to maximize the worst-case guarantee.

\begin{proposition}[Optimal orientation guarantee]
\label{prop:robust_orientation_mounting}
For $0<\Delta\le\pi$, the cone parameter maximizing $\lambda_\Delta(\mu)$ over $\mu\in[0,1]$ and the corresponding guarantee are
\begin{equation}
    \mu^\star(\Delta)
    =\frac{U_\Delta+S_\Delta}
    {1+U_\Delta+2S_\Delta},~~
    \lambda_\Delta^\star
    =\frac{U_\Delta-S_\Delta^2}
    {1+U_\Delta+2S_\Delta}.
\label{eq:complex_mu_opt}
\end{equation}
In particular,
$\mu^\star(\pi)=\lambda_\pi^\star=1/3$.

For a severely restricted swivel interval,
\begin{equation*}
    \mu^\star(\Delta)
    =\frac12-\frac{\Delta^2}{24}
    +O(\Delta^4),~~
    \lambda_\Delta^\star
    =\frac{\Delta^4}{180}
    +O(\Delta^6).
\end{equation*}
\end{proposition}
\begin{proof}
    See the supplementary material.
\end{proof}

Proposition~\ref{prop:robust_orientation_mounting} provides a simple geometric interpretation of robust polarization reconfiguration. With a full swivel revolution, the optimal value $\mu=1/3$ balances the axial and transverse orientation coverage, giving $\bm Q_\pi=\bm I/3$. For a narrow feasible interval, the optimal cone parameter approaches $1/2$, but the maximum guarantee decreases as $O(\Delta^4)$. Thus, even an optimally shaped orientation cone offers limited worst-case polarization reconfiguration when the prescribed task leaves very little self-motion.

The preceding guarantee applies to arbitrary complex effective fields. When the effective field is linearly polarized, a stronger result is possible: we can characterize exactly when the feasible dipole orientations achieve perfect polarization matching.

Suppose $\bm f=a\bm\epsilon$, where $a\in\mathbb C\setminus\{0\}$ and $\bm\epsilon\in\mathbb R^3$ is a real unit polarization direction. Then $\frac{P(\tilde{\psi})}{\|\bm f\|^2}
    =[\bm n(\tilde{\psi})^{\mathsf T}
    \bm\epsilon]^2$.
Because $\bm\epsilon$ and $-\bm\epsilon$ represent the same linear polarization, choose the representative satisfying $\bm e^{\mathsf T}\bm\epsilon\ge0$ and write
\begin{equation*}
    \bm\epsilon=
    \begin{bmatrix}
        \sqrt{1-\eta}\cos\varphi_f\\
        \sqrt{1-\eta}\sin\varphi_f\\
        \sqrt{\eta}
    \end{bmatrix},
    \qquad
    \eta=|\bm e^{\mathsf T}\bm\epsilon|^2.
\end{equation*}
Here, $\eta$ is the axial power fraction of the field polarization, while $\varphi_f$ is the azimuth of its transverse component in the chosen basis. If $\eta=1$, set $\varphi_f=0$, since the transverse azimuth is immaterial.

The polarization projection becomes $\bm n(\tilde{\psi})^{\mathsf T}\bm\epsilon
    =\sqrt{\mu\eta} +\sqrt{(1-\mu)(1-\eta)}
    \cos(\tilde{\psi}-\varphi_f)$.
For a finite feasible interval, the achievable alignment depends on how close the available swivel angles can approach the desired polarization direction. We use the following circular distance to express this dependence.

\begin{definition}[Circular distance]
\label{def:circular_distance}
For $x\in\mathbb R$ and a nonempty compact set $\mathcal S\subset\mathbb R$, define
\begin{equation*}
    d_{\mathbb T}(x,\mathcal S)
    \triangleq
    \min_{\substack{y\in\mathcal S\\z\in\mathbb Z}}
    |x-y+2\pi z|
    \in[0,\pi].
\end{equation*}
This is the minimum angular separation between $x$ and the set $\mathcal S$, accounting for the $2\pi$ periodicity of angles. The subscript $\mathbb T$ denotes angles identified modulo $2\pi$.
\end{definition}

Let $d_0
    \triangleq d_{\mathbb T}
    (\varphi_f,[-\Delta,\Delta])$ and $d_\pi
    \triangleq d_{\mathbb T}
    (\varphi_f+\pi,[-\Delta,\Delta])$.
These distances quantify how closely the feasible swivel interval approaches the two opposite transverse polarization directions.

\begin{corollary}[Linear polarization matching]
\label{cor:specified_linear_field}
For the linearly polarized effective field above,
\begin{equation*}
\begin{aligned}
    \frac{P^\star}{\|\bm f\|^2}
    =\max\Bigl\{&
    \bigl[\sqrt{\mu\eta}
    +\sqrt{(1-\mu)(1-\eta)}
    \cos d_0\bigr]^2,\\
    &
    \bigl[\sqrt{\mu\eta}
    -\sqrt{(1-\mu)(1-\eta)}
    \cos d_\pi\bigr]^2
    \Bigr\}.
\end{aligned}
\end{equation*}
Moreover, $P^\star=\|\bm f\|^2$ if and only if the feasible orientation range contains a dipole direction parallel or antiparallel to $\bm\epsilon$.

Over a full $2\pi$ swivel range,
\begin{equation*}
\begin{aligned}
    \frac{P^\star}{\|\bm f\|^2}
    &=\left[
    \sqrt{\mu\eta}
    +\sqrt{(1-\mu)(1-\eta)}
    \right]^2\\
    &\le1,
\end{aligned}
\end{equation*}
with equality if and only if $\mu=\eta$.
\end{corollary}
\begin{proof}
    See the supplementary material.
\end{proof}

Corollary~\ref{cor:specified_linear_field} reveals the geometric condition for perfect polarization matching. Over a complete swivel revolution, the dipole can align with a linearly polarized effective field precisely when its orientation cone has the same axial power fraction as that field, i.e., $\mu=\eta$. Over a restricted swivel interval, this condition alone is insufficient: the required direction must also lie on the accessible portion of the cone.

Therefore, the finite task-preserving motion limits polarization matching in two ways. The mounting and task determine the cone geometry through $\mu$, while the feasible swivel interval determines which directions on that cone can actually be reached.

\subsection{Multipath Phase Control through Translation}

To isolate the effect of antenna translation, hold the nonzero antenna responses $g_\ell$ fixed over the feasible interval while retaining the circular phase-center motion. Write $\alpha_\ell g_\ell=A_\ell e^{j\chi_\ell}$, $A_\ell>0$, and define $\chi_{\ell i}\triangleq\chi_\ell-\chi_i$. In the centered swivel coordinate, the received power is
\begin{equation*}
    P(\tilde{\psi})
    =\sum_{\ell=1}^{L}A_\ell^2+2\sum_{\ell<i}A_\ell A_i
    \cos\bigl(
    \chi_{\ell i}
    -\vartheta_{\ell i}(\tilde{\psi})
    \bigr),
\end{equation*}
where $\vartheta_{\ell i}$ is the relative propagation phase induced by translation.

The ability to improve multipath combining depends on how much relative phase the feasible antenna motion can generate. Define the projected directional separation between paths $\ell$ and $i$ as $\xi_{\ell i}
    \triangleq\left\|
    (\bm I-\bm e\bm e^{\mathsf T})
    (\bm u_\ell-\bm u_i)
    \right\|$,
so that $r_{\ell i}=k\rho\xi_{\ell i}$. A larger $\xi_{\ell i}$ yields a larger relative phase variation amplitude for the same motion radius.

For $\xi_{\ell i}>0$, define
\begin{equation*}
    \gamma_{\ell i}
    \triangleq\operatorname{atan2}
    \bigl(
    (\bm u_\ell-\bm u_i)^{\mathsf T}\bm e_2,
    (\bm u_\ell-\bm u_i)^{\mathsf T}\bm e_1
    \bigr)
    -\delta_p-\psi_c.
\end{equation*}
Its value is immaterial when $\xi_{\ell i}=0$. The parameter $\gamma_{\ell i}$ describes the angular alignment between the projected path-direction difference and the center of the feasible antenna arc. The relative phase variation is then $\vartheta_{\ell i}(\tilde{\psi})
    =r_{\ell i}\cos(
    \tilde{\psi}-\gamma_{\ell i})$.
Thus, the attainable phase variation depends on the motion radius $\rho$, projected path separation $\xi_{\ell i}$, and arc alignment $\gamma_{\ell i}$.

To determine the attainable relative-phase range, define $d_{\ell i}^{+}
    \triangleq d_{\mathbb T}
    (\gamma_{\ell i},[-\Delta,\Delta])$
    and
    $d_{\ell i}^{-}
    \triangleq d_{\mathbb T}
    (\gamma_{\ell i}+\pi,[-\Delta,\Delta])$.
The first distance determines how closely the feasible arc approaches the maximum of the relative-phase sinusoid, while the second determines how closely it approaches the minimum.

The attainable relative-phase interval is $\Theta_{\ell i}
    =
    \big[
    -r_{\ell i}\cos d_{\ell i}^{-},
   \allowbreak
    r_{\ell i}\cos d_{\ell i}^{+}
    \big]$
with width $W_{\ell i}
    =\allowbreak r_{\ell i}
    \big(
    \cos d_{\ell i}^{+}
    \allowbreak +\cos d_{\ell i}^{-}
    \big)$.
For a full $2\pi$ swivel range, the entire relative-phase sinusoid is accessible, giving $\Theta_{\ell i}=[-r_{\ell i},r_{\ell i}]$ and $W_{\ell i}=2r_{\ell i}$.
The dependence on the interval alignment $\gamma_{\ell i}$ then disappears.

The two-path case admits an exact characterization of how much coherent combining can be achieved through feasible antenna translation.

\begin{proposition}[Two-path phase coverage]
\label{prop:two_path_translation}
Consider two nonzero paths with constant antenna responses. Then
\begin{equation}
 P^\star
 =A_1^2+A_2^2
 +2A_1A_2\cos d_{\mathbb T}(\chi_{12},\Theta_{12}).
 \label{eq:two_path_exact_power}
\end{equation}
Perfect coherent combining is possible for the given $\chi_{12}$ if and only if $d_{\mathbb T}(\chi_{12},\Theta_{12})=0$, and it is possible for every phase offset if and only if $W_{12}\ge2\pi$. For a full $2\pi$ swivel range, universal phase coverage is possible only if $\xi_{12}>0$, in which case the condition reduces to
\begin{equation}
 \rho\ge\frac{\lambda_c}{2\xi_{12}}.
 \label{eq:full_circle_phase_radius}
\end{equation}
\end{proposition}
\begin{proof}
    See the supplementary material.
\end{proof}

For two paths, perfect combining requires the antenna motion to compensate their initial phase difference modulo $2\pi$. If the attainable relative-phase interval spans at least $2\pi$, a suitable swivel angle exists for every initial phase offset.

Importantly, the phase coverage condition depends on the feasible arc, not only on the nominal circle radius. Over a full revolution, the condition simplifies to a minimum radius determined by the carrier wavelength and projected path separation. Over a restricted interval, the attainable width $W_{12}$ also depends on $\Delta$ and $\gamma_{12}$. Hence, a large phase-center circle does not necessarily provide sufficient phase coverage when the physical task permits only a short or unfavorably aligned portion of that circle.

\begin{remark}[Multipath phase limitation]
\label{rem:multipath_phase_limit}
For $L\ge3$ active paths with constant antenna responses, coherent combining requires controlling multiple relative phases. Define
\begin{equation*}
 \bm\vartheta(\tilde{\psi})
 =[\vartheta_{21}(\tilde{\psi}),\ldots,\vartheta_{L1}(\tilde{\psi})]^{\mathsf T}
 \pmod{2\pi}
\end{equation*}
Although this vector contains $L-1$ relative phases, all of them are determined by the same scalar swivel angle. As $\tilde{\psi}$ varies, the vector therefore traces a one-dimensional analytic curve in the $(L-1)$-dimensional relative-phase space. In general, the relative phases cannot be adjusted independently to align all paths simultaneously. In particular, individual phase coverage for every path pair does not guarantee simultaneous coherent combining of all paths.
\end{remark}

These results isolate the phase-reconfiguration capability of embodied antenna translation. In the complete channel, however, the same swivel angle simultaneously changes the dipole orientation.
Consequently, the isolated orientation and translation results provide physical insight, while actual configuration selection must use the coupled response in~\eqref{eq:coupled_channel_swivel}.

To conclude this section, we have identified three factors governing the wireless value of task-preserving self-motion: the antenna pose variation induced by the host, the feasible portion of that motion under the prescribed task, and the propagation environment with which the motion interacts. Because a fixed physical mounting can produce different antenna motions and feasible responses under different tasks, mounting design should account for the complete coupled received power rather than optimizing translation or orientation in isolation. This motivates the cross-task mounting design in Section~\ref{sec:optimization}.

\section{Cross-Task Antenna Mounting Design}
\label{sec:optimization}

The physical mounting must be selected before operation, while the feasible swivel angle can be adapted to each prescribed task and channel realization. Changing the task alters robot quantities such as $r_E$, $c_X$, and $s_X$, so the same mounting can produce different antenna motion trajectories. Although the mounting controls the circle radius $\rho$ and orientation parameter $\mu$, these descriptors alone do not determine the complete coupled wireless response. Their exact attainable ranges and related orientation design conditions are derived in the supplementary material. We therefore design one physical mounting across representative tasks and propagation conditions using the complete received-power model.

\subsection{Cross-Task Mounting Objective}

Let $\mathcal M\subseteq \{U,F\}\times[0,1]\times[0,2\pi]\times[0,\pi]$ be a nonempty compact set of admissible mountings. This set can exclude inaccessible link regions and impose restrictions associated with antenna clearance, mechanical access, or cable routing. For each supporting link $X\in\{U,F\}$, define $\mathcal M_X=\{(\tau,\phi,\beta): (X,\tau,\phi,\beta)\in\mathcal M\}$.

Consider $N_T$ representative tasks
$\{\mathcal T_a\}_{a=1}^{N_T}$, each prescribing a fixed full end-effector pose. Task $\mathcal T_a$ determines a feasible swivel interval $\Psi_a$ and, together with a mounting $\bm m$, the antenna motion parameters in Proposition~\ref{prop:srs_pose}. In particular, let $\rho_a(\bm m)$ and $\mu_a(\bm m)$ denote the resulting phase-center radius and orientation-cone parameter.

As assumed in Section~\ref{sec:model}, the conformal antenna has negligible thickness in the robot collision model, and its mounting does not further restrict the self-motion. Thus, $\Psi_a$ depends on the prescribed task and robot constraints but is independent of $\bm m\in\mathcal M$.

For task $\mathcal T_a$, let $\mathcal H_a=\{\bar\alpha_{a,\ell},\bm u_{a,\ell},\bm\epsilon_{a,\ell}\}_{\ell=1}^{L}$ denote the channel realization, and let $P_a(\psi;\bm m,\mathcal H_a)$ be the received power obtained from the coupled channel model in Section~\ref{sec:response}. Following Definition~\ref{def:feasible_power_variation}, the power available after selecting the best feasible configuration is
\begin{equation}
  P_a^\star(\bm m,\mathcal H_a)
  \triangleq
  \max_{\psi\in\Psi_a}
  P_a(\psi;\bm m,\mathcal H_a).
  \label{eq:configuration_optimization}
\end{equation}
Let $\pi_a\ge0$ be the long-term weight of task $\mathcal T_a$, with $\sum_{a=1}^{N_T}\pi_a=1$.

For a fixed physical mounting $\bm m$, define its cross-task mounting utility as the weighted average received power after feasible swivel selection:
\begin{equation}
  J(\bm m)
  \triangleq
  \sum_{a=1}^{N_T}
  \pi_a
  \mathbb E_{\mathcal H_a}
  \left[
    P_a^\star(\bm m,\mathcal H_a)
  \right].
  \label{eq:mounting_statistical_objective}
\end{equation}

The task weights may represent the fraction of long-term operation spent on each task or application-specific design priorities. The channel distributions can differ across tasks, for example, when the robot performs them at different operating locations or under different propagation conditions. The objective accounts for these variations while requiring the same physical mounting to be used throughout.

The cross-task mounting design problem is
\begin{equation}
  \bm m^\star
  \in\arg\max_{\bm m\in\mathcal M}J(\bm m).
  \label{eq:joint_mounting_problem}
\end{equation}
Here, the mounting determines an entire set of task-dependent antenna poses rather than a single antenna position. The optimization therefore chooses the physical embodiment through which the host's residual motion can be exploited for communication.
The expectation in~\eqref{eq:mounting_statistical_objective} is taken after maximizing over $\psi$, allowing different channel realizations to favor different configurations.
Under the finite-moment condition detailed in the supplementary material, $J(\bm m)$ is continuous on the compact set $\mathcal M$, and an optimal mounting exists.

\subsection{Approximating the Mounting Objective}

The objective in~\eqref{eq:mounting_statistical_objective} involves an expectation over channel realizations and an inner maximization over the feasible swivel interval. We approximate these two operations separately: first using channel samples to evaluate the expectation, and then using a finite swivel grid to evaluate the inner maximization.

For each task $a$, let
$\{\mathcal H_{a,r}\}_{r=1}^{N_c}$
be independent and identically distributed realizations of $\mathcal H_a$. The sample average approximation (SAA)~\cite{shapiro_stochastic_2014} of the mounting utility is
\begin{equation}
  \widehat J_{N_c}(\bm m)
  \triangleq
  \sum_{a=1}^{N_T}
  \frac{\pi_a}{N_c}
  \sum_{r=1}^{N_c}
  P_a^\star(\bm m,\mathcal H_{a,r}).
  \label{eq:saa_mounting_objective}
\end{equation}
For finite $N_c$, evaluating~\eqref{eq:saa_mounting_objective} still requires maximizing the received power over a continuous swivel interval for every channel sample. Because the inner problem is one dimensional, it can be approximated using a finite grid.

For each task $a$, let
$\Psi_{a,Q}\subseteq\Psi_a$
be a finite swivel grid containing both interval endpoints, with $Q$ indexing the grid resolution, and let $\delta_{\psi,a}$ denote the largest spacing between adjacent grid points. Define the grid-selected power as
\begin{equation*}
  \widetilde P_{a,Q}^\star
  (\bm m,\mathcal H_a)
  \triangleq
  \max_{\psi\in\Psi_{a,Q}}
  P_a(\psi;\bm m,\mathcal H_a).
\end{equation*}
The corresponding grid-based SAA objective is
\begin{equation}
  \widetilde J_{N_c,Q}(\bm m)
  \triangleq
  \sum_{a=1}^{N_T}
  \frac{\pi_a}{N_c}
  \sum_{r=1}^{N_c}
  \widetilde P_{a,Q}^\star
  (\bm m,\mathcal H_{a,r}).
  \label{eq:grid_saa_objective}
\end{equation}

\begin{algorithm}[!t]
\caption{Offline Cross-Task Antenna Mounting Design}
\label{alg:offline_mounting}
\begin{algorithmic}[1]
\Require Admissible mounting set $\mathcal M$, task weights
$\{\pi_a\}$, channel samples $\{\mathcal H_{a,r}\}$,
and swivel grids $\{\Psi_{a,Q}\}$
\For{$X\in\{U,F\}$ with $\mathcal M_X\neq\emptyset$}
  \State Apply SHGO to minimize
  $-\widetilde J_{N_c,Q}(\bm m)$ over
  $(\tau,\phi,\beta)\in\mathcal M_X$
  \State For each objective evaluation, compute
  $\widetilde J_{N_c,Q}(\bm m)$ using
  $\max_{\psi\in\Psi_{a,Q}}
  P_a(\psi;\bm m,\mathcal H_{a,r})$
  for all tasks $a$ and samples $r$
  \State Store the best mounting candidate
  $\widetilde{\bm m}_X$
  and its objective value
\EndFor
\State \Return the candidate with the larger objective value
\end{algorithmic}
\end{algorithm}

The loss from restricting swivel selection to a finite grid, including its effect on the mounting optimization, is bounded in the supplementary material (Proposition~\ref{prop:swivel_grid_approximation}).

\subsection{Offline Optimization and Online Adaptation}

After the sample and grid approximations, the remaining problem is to maximize $\widetilde J_{N_c,Q}(\bm m)$ over the admissible physical mountings. Although the objective is generally nonconvex, the continuous mounting search has only three variables $(\tau,\phi,\beta)$ for each supporting link $X\in\{U,F\}$. The number of representative tasks affects the cost of evaluating the objective, but not the dimension of this search.

The nonconvexity arises because the antenna translation phases and polarization responses vary jointly with the mounting, and the maximizing swivel angle can differ across tasks and channel realizations. Since the mounting is chosen offline, we use a derivative-free global optimization method.

Specifically, we employ simplicial homology global optimization (SHGO)~\cite{endres_shgo_2018} to minimize $-\widetilde J_{N_c,Q}$ for each supporting link. Each objective evaluation computes the best grid-selected power over the channel samples and feasible swivel grid of every representative task. The better mounting from the two link searches is retained. The procedure is summarized in Algorithm~\ref{alg:offline_mounting}.

\textit{Computational complexity.} For a fixed mounting candidate, the received-power evaluations for different tasks, channel samples, and swivel grid points are independent and can be computed in parallel. Let $N_{\psi,a}=|\Psi_{a,Q}|$ denote the number of swivel grid points for task $a$. With $N_c$ channel samples per task and $L$ paths, evaluating the grid-based objective requires $\mathcal O\!\left(N_cL\sum_{a=1}^{N_T}N_{\psi,a}\right)$ operations. Let $N_{\rm eval}$ denote the number of objective evaluations per link search. Since there are at most two candidate links, the total computational order is $\mathcal O\!\left(N_{\rm eval}N_cL\sum_{a=1}^{N_T}N_{\psi,a}\right)$.

After installation, the mounting $\bm m^\star$ is fixed. For the current task $\mathcal T_a$ and channel realization $\mathcal H_a$, communication only needs to solve the scalar configuration-selection problem in~\eqref{eq:configuration_optimization}. If a swivel grid is used online, the computational order is $\mathcal O(LN_{\psi,a})$ per channel realization. No reinstallation or modification of the prescribed physical task is required.

\section{Numerical Results}
\label{sec:numerical}

\begin{figure}[!t]
    \centering
    \includegraphics[width=0.8\columnwidth]{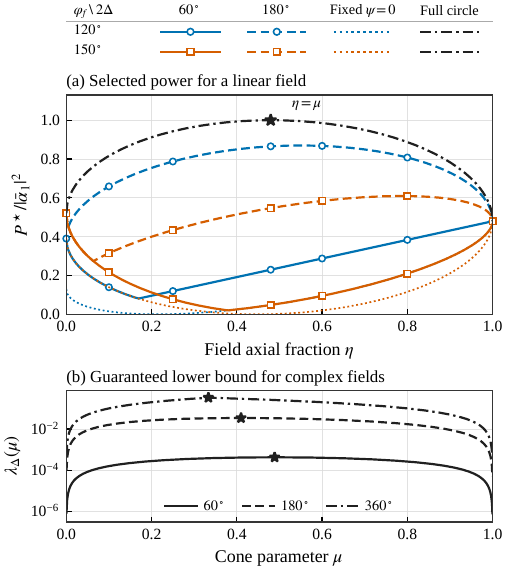}
    \caption{Task-constrained polarization reconfiguration. (a) Normalized selected power for a linearly polarized field under $\mathcal T_2$ and $\bm m_0$, with different field azimuths and feasible swivel spans. The dotted curves correspond to the fixed configuration $\psi=0$, and the star marks perfect matching at $\eta=\mu$ over a full swivel circle. (b) Orientation guarantee $\lambda_\Delta(\mu)$ for arbitrary complex effective fields. Stars indicate the optimal cone parameters $\mu^\star(\Delta)$.}
    \label{fig:orientation_capability}
\end{figure}

\begin{figure}[!t]
 \centering
 \includegraphics[width=0.8\columnwidth]{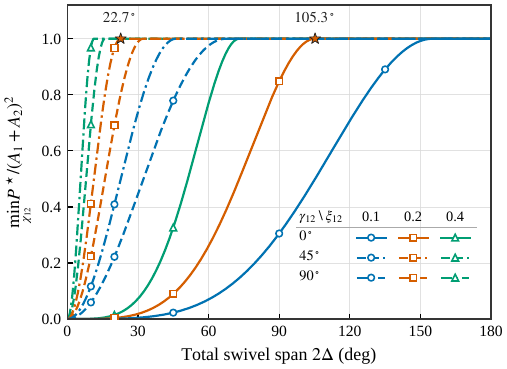}
 \caption{Translation capability for two equal effective path amplitudes with fixed antenna responses. The ordinate is the normalized worst selected power $\min_{\chi_{12}}P^\star/(A_1+A_2)^2$. Curves vary the projected path separation $\xi_{12}$ and finite-arc alignment $\gamma_{12}$. Stars highlight the universal coverage thresholds for $\xi_{12}=0.2$ at $\gamma_{12}=0^\circ$ and $90^\circ$.}
 \label{fig:translation_capability}
\end{figure}

\begin{figure}[!t]
 \centering
 \includegraphics[width=0.8\columnwidth]{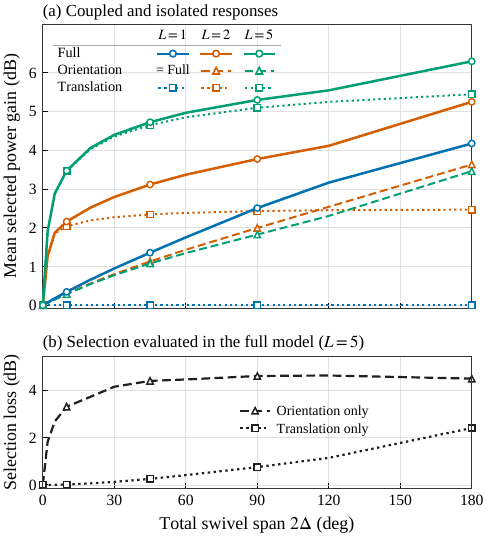}
 \caption{Coupled reconfigurability under $\mathcal T_2$ and $\bm m_0$. (a) Mean selected power gain for the full, orientation, and translation responses with different numbers of active paths. (b) Selected power loss from using an isolated-response swivel instead of the full-response optimum for $L=5$.}
 \label{fig:coupled_reconfigurability}
\end{figure}

This section illustrates the orientation and translation capabilities identified in Section~\ref{sec:reconfigurability}, evaluates their coupled effect on swivel selection, and examines the fixed mounting design in Section~\ref{sec:optimization}. Unless stated otherwise, $L_1=0.32$ m, $L_2=0.28$ m, $\varrho_U=0.03$ m, $\varrho_F=0.025$ m, and the carrier frequency is $28$ GHz. Figs.~\ref{fig:orientation_capability}--\ref{fig:coupled_reconfigurability} use the representative task $\mathcal T_2$, whose shoulder-to-wrist axis has $d=0.38$ m, azimuth $-20^\circ$, and elevation $-20^\circ$, together with the reference mounting $\bm m_0=(U,0.5,0,0)$. This gives $\rho/\lambda_c=12.72$ and $\mu=0.479$.

For the multipath experiments, five nominal angular clusters have azimuths $175^\circ$, $210^\circ$, $135^\circ$, $250^\circ$, and $100^\circ$, with elevations $0^\circ$, $15^\circ$, $-10^\circ$, $25^\circ$, and $35^\circ$, respectively. Their sampling probabilities are proportional to the linear powers corresponding to $0$, $-2.5$, $-5$, $-7$, and $-9$ dB, and their nominal linear polarization angles are $10^\circ$, $55^\circ$, $95^\circ$, $140^\circ$, and $30^\circ$. Conditional on the sampled clusters, the paths have equal coefficient magnitudes $|\bar\alpha_\ell|=1/\sqrt{L}$ and independent uniform phases. Their directions and polarization angles are perturbed about the corresponding cluster centers with standard deviations $4^\circ$ and $10^\circ$, respectively.

\subsection{Orientation and Translation Capabilities}

Fig.~\ref{fig:orientation_capability} isolates the orientation mechanism under $\mathcal T_2$ and $\bm m_0$. For a single linearly polarized path, translation contributes only a unit-modulus phase, so the selected power is governed by the dipole orientation. Fig.~\ref{fig:orientation_capability}(a) varies the field axial fraction $\eta$ for two field azimuths and several feasible swivel spans. Over a full swivel circle, Corollary~\ref{cor:specified_linear_field} predicts perfect polarization matching at $\eta=\mu$. Restricting the swivel interval lowers the attainable power and introduces a dependence on the field azimuth relative to the available arc. At $\eta\approx\mu$, for example, a $60^\circ$ total span gives normalized selected powers of about $0.23$ and $0.05$ for the two azimuths, whereas the full circle reaches unity.

Fig.~\ref{fig:orientation_capability}(b) shows the guarantee in Theorem~\ref{thm:orientation_coverage}. Here, $\mu$ is varied to examine different orientation cones, rather than fixed at the value $\mu=0.479$ induced by $\bm m_0$. The optimized guarantee increases from $4.20\times10^{-4}$ for a $60^\circ$ total span to $1/3$ for a full circle, while its maximizing orientation parameter moves from $0.489$ to $1/3$. The finite-arc curves therefore quantify how the guaranteed orientation capability contracts as the prescribed task leaves less self-motion. For the full circle, the same quantity becomes the exact worst-case value in~\eqref{eq:full_circle_complex_capability}.

To isolate translation, we fix the antenna responses as $g_\ell(\psi)=g_\ell(0)$ under the same task and mounting, and use two paths with equal effective amplitudes. Fig.~\ref{fig:translation_capability} plots the minimum normalized selected power over all initial two-path phase offsets. A value of one means that the available swivel arc provides universal phase coverage.

The curves follow Proposition~\ref{prop:two_path_translation}: the attainable phase width is controlled jointly by $r_{12}=k\rho\xi_{12}$ and $\gamma_{12}$. With $\xi_{12}=0.2$, universal coverage requires about $22.7^\circ$ of total swivel for $\gamma_{12}=90^\circ$, but about $105.3^\circ$ for $\gamma_{12}=0^\circ$. Increasing $\xi_{12}$ reduces the required span because the same phase center motion produces a larger relative phase excursion. The large separation between these thresholds also shows why the phase center circle radius alone does not determine finite-arc translation capability.

\subsection{Coupled Reconfigurability}

The full response couples orientation and translation through the same swivel angle. For comparison, we also evaluate an orientation response obtained by fixing $\bm p_A$ at its value for $\psi=0$, and a translation response obtained by fixing $\bm n$ at its value for $\psi=0$. These isolated responses serve only to distinguish the two wireless mechanisms and are not independently controllable antenna motions. All three responses coincide at $\psi=0$. Their selected power gains are computed from the ratio between the mean selected power and the mean power at $\psi=0$ and are reported in dB. The results use $10^4$ channel realizations and a $0.03125^\circ$ swivel grid.

\begin{figure}[!t]
 \centering
 \includegraphics[width=0.72\columnwidth]{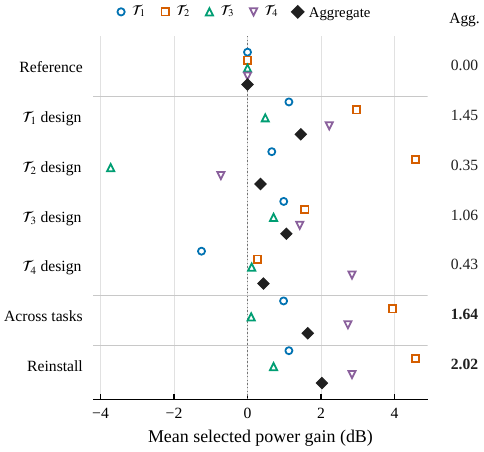}
 \caption{Mounting design across four prescribed tasks. Each row corresponds to one mounting strategy. Colored markers show per-task gain relative to the reference mounting $\bm m_0$ with its own optimal swivel selection. Diamonds show the aggregate gain over the four equally weighted tasks, and the values on the right report the same aggregate gain in dB. ``Reinstall'' uses the mounting optimized separately for each task.}
 \label{fig:mounting_across_tasks}
\end{figure}

Fig.~\ref{fig:coupled_reconfigurability}(a) first recovers the single-path case: the full and orientation responses coincide for $L=1$, while translation alone gives no power gain. Multipath activates the relative phase mechanism. At a $90^\circ$ total span with $L=5$, the full selected power gain is $5.30$ dB, compared with $1.83$ dB from orientation alone and $5.10$ dB from translation alone. The full gain continues to increase to $6.30$ dB at a $180^\circ$ span, showing that the wider self-motion interval still exposes useful configurations after much of the early translation gain has already been obtained.

Fig.~\ref{fig:coupled_reconfigurability}(b) evaluates in the full model the swivel choices obtained from the isolated responses. For $L=5$ and a $90^\circ$ span, choosing the swivel from the orientation response loses $4.61$ dB relative to selection using the full response, while selection based on translation loses $0.75$ dB. The loss from selection based on translation remains below $0.07$ dB up to a $20^\circ$ span and grows as the interval widens. For this trajectory at $28$ GHz, translation therefore supplies most of the selection gain over a short arc, while orientation increasingly changes which configuration is best over wider self-motion ranges.

\subsection{Mounting Design Across Tasks}

We finally evaluate one fixed mounting over four prescribed tasks. For $\mathcal T_1$--$\mathcal T_4$, the shoulder-to-wrist distances are $0.28$, $0.38$, $0.47$, and $0.54$ m, the axis azimuths are $-60^\circ$, $-20^\circ$, $25^\circ$, and $65^\circ$, and the elevations are $30^\circ$, $-20^\circ$, $25^\circ$, and $-10^\circ$, respectively. The tasks have equal weights, share the feasible swivel interval $[-45^\circ,45^\circ]$, and use the same multipath statistics. The admissible mounting set includes both links with $\tau\in[0.1,0.9]$, $\phi\in[0,2\pi]$, and $\beta\in[0,\pi]$. The mounting design uses $N_c=256$ independently sampled channels per task and a 721-point swivel grid. Reported powers are evaluated using $10^4$ independent test channels per task and a 1441-point grid.

The strategies in Fig.~\ref{fig:mounting_across_tasks} include the reference mounting, four mountings optimized separately for $\mathcal T_1$--$\mathcal T_4$, one mounting optimized across all four tasks, and a reinstallation benchmark that uses the corresponding task-specific mounting. The gain on each task is measured relative to the mean selected power of $\bm m_0$ for the same task. Unlike the fixed $\psi=0$ reference in Fig.~\ref{fig:coupled_reconfigurability}, $\bm m_0$ here also uses its best feasible swivel angle for each channel realization. The aggregate gain is the ratio of the task-weighted mean selected powers, converted to dB after averaging in the linear power domain.

A mounting optimized for one task can perform poorly when the prescribed task changes. The $\mathcal T_2$ design, for example, gains $4.57$ dB on $\mathcal T_2$ but loses $3.72$ dB on $\mathcal T_3$. The design across tasks gives an aggregate gain of $1.64$ dB, compared with $1.45$ dB for the best fixed mounting optimized for one task and $2.02$ dB for task-specific reinstallation. It therefore captures most of the reinstallation benefit with one physical mounting. Its gain on $\mathcal T_3$ is only $0.10$ dB, illustrating that maximizing weighted-average power need not yield comparable gains on every task. It selects the upper arm with $(\tau,\phi,\beta)=(0.894,54.7^\circ,138.0^\circ)$, illustrating how the long-term objective in~\eqref{eq:mounting_statistical_objective} balances the antenna motions generated under the different prescribed tasks.

\section{Conclusion}
This paper established embodied MAs as a new MA architecture and provided a theoretical foundation for exploiting task-preserving motion as a wireless reconfiguration resource.
Its broader significance lies in extending the design of antenna mobility beyond dedicated actuation to the physical embodiment of intelligent agents.
Our analysis reveals a fundamental distinction: mechanical freedom does not necessarily translate into useful wireless reconfigurability.
Rather, the attainable communication benefit depends on which antenna configurations remain feasible under the physical task and how their coupled translation and orientation interact with the propagation environment.
This insight shifts the design perspective from simply exploiting available mechanical freedom to understanding how that freedom can be made valuable for wireless communication.
More broadly, it points toward a new frontier in which the physical embodiment and wireless capabilities of intelligent agents are designed together, making how a machine is built an integral part of how it communicates.

\bibliographystyle{IEEEtran}
\bibliography{references}
\clearpage
\section*{Supplementary Material}

\newcommand{\suppheading}[1]{%
  \par\addvspace{1.0\baselineskip}%
  \noindent\textbf{#1}\par\nobreak\smallskip\noindent%
}
\newcommand{\suppsubheading}[1]{%
  \par\addvspace{0.7\baselineskip}%
  \noindent\textit{#1}\par\nobreak\smallskip\noindent%
}
\newenvironment{suppproof}[1]{%
  \par\addvspace{0.55\baselineskip}%
  \noindent\textit{Proof of #1.}\enspace\ignorespaces%
}{%
  \unskip\nobreak\hfill\qedsymbol\par%
}

\newtheorem{suppproposition}{Proposition}
\renewcommand{\thesuppproposition}{S\arabic{suppproposition}}

\suppheading{Proofs of the Main Results}

\begin{suppproof}{Proposition~\ref{prop:srs_pose}: SRS antenna motion}
Substituting the mounting expressions into the first line of~\eqref{eq:surface_mounting_pose} gives
\[
\begin{aligned}
 \bm p_A=\bm p_c
 &+\bigl(\rho_{0,X}+\varrho_Xc_X\cos\phi\bigr)\bm e_r(\psi)\\
 &+\varrho_X\sin\phi\,\bm e_t(\psi).
\end{aligned}
\]
The two transverse coefficients have magnitude and polar angle
\[
\begin{aligned}
 \rho&=\sqrt{\bigl(\rho_{0,X}+\varrho_Xc_X\cos\phi\bigr)^2
 +\varrho_X^2\sin^2\phi},\\
 \delta_p&=\operatorname{atan2}\!\left(
 \varrho_X\sin\phi,\rho_{0,X}+\varrho_Xc_X\cos\phi\right),
\end{aligned}
\]
which gives~\eqref{eq:srs_position_locus} and~\eqref{eq:surface_position_descriptors}. Similarly, the second line of~\eqref{eq:surface_mounting_pose} becomes
\[
\begin{aligned}
 \bm n=\nu\bm e
 &+\bigl(\sigma_Xs_X\cos\beta-c_X\sin\phi\sin\beta\bigr)\bm e_r(\psi)\\
 &+\cos\phi\sin\beta\,\bm e_t(\psi),
\end{aligned}
\]
where \(\nu=c_X\cos\beta+\sigma_Xs_X\sin\phi\sin\beta\). Since \(\|\bm n\|=1\), the transverse coefficient vector has norm \(\sqrt{1-\nu^2}\). Its polar angle is exactly \(\delta_n\) in~\eqref{eq:surface_orientation_descriptors}, yielding~\eqref{eq:srs_orientation_motion}. The conventions for vanishing transverse components follow because the corresponding angular offset is then immaterial.
\end{suppproof}

\begin{suppproof}{Theorem~\ref{thm:coupled_fourier_representation}: Fourier representation}
For each path, the Jacobi-Anger expansion gives
\[
 e^{-j\kappa_\ell\cos(\psi+\delta_p-\zeta_\ell)}
 =\sum_{r=-\infty}^{\infty}(-j)^rJ_r(\kappa_\ell)
 e^{jr(\psi+\delta_p-\zeta_\ell)}.
\]
For finite \(\kappa_\ell\), this series is absolutely convergent. Multiplying it by \(g_\ell(\psi)=\sum_{s=-1}^{1}o_{\ell,s}e^{js\psi}\) gives
\[
\begin{aligned}
 \alpha_\ell g_\ell(\psi)
 e^{-j\kappa_\ell\cos(\psi+\delta_p-\zeta_\ell)}
 ={}&\alpha_\ell
 \sum_{s=-1}^{1}\sum_{r=-\infty}^{\infty}
 o_{\ell,s}(-j)^rJ_r(\kappa_\ell)\\
 &\times e^{jr(\delta_p-\zeta_\ell)}e^{j(r+s)\psi}.
\end{aligned}
\]
Setting \(n=r+s\) and summing over the finite path set yields
\[
 H_n=\sum_{\ell=1}^{L}\alpha_\ell
 \sum_{s=-1}^{1}o_{\ell,s}(-j)^{n-s}J_{n-s}(\kappa_\ell)
 e^{j(n-s)(\delta_p-\zeta_\ell)},
\]
which is~\eqref{eq:channel_fourier_coefficients}. Absolute convergence is preserved by the finite sum over \(\ell\) and the three-term sum over \(s\), so \(\sum_n|H_n|<\infty\).

The product of the two absolutely convergent channel series can therefore be rearranged as a Cauchy product:
\[
\begin{aligned}
 P(\psi)
 &=\left(\sum_nH_ne^{jn\psi}\right)
   \left(\sum_qH_q^*e^{-jq\psi}\right)\\
 &=\sum_{m=-\infty}^{\infty}
 \left(\sum_{n=-\infty}^{\infty}H_{n+m}H_n^*\right)e^{jm\psi}.
\end{aligned}
\]
This proves~\eqref{eq:power_fourier_series}. Moreover,
\(
\sum_m|C_m|\le(\sum_n|H_n|)^2<\infty
\), so the power series is also absolutely convergent. Since \(P(\psi)\) is real, its Fourier coefficients satisfy \(C_{-m}=C_m^*\).

If \(\rho=0\), then every \(\kappa_\ell=0\). Because \(J_0(0)=1\) and \(J_r(0)=0\) for \(r\neq0\), only \(n=s\in\{-1,0,1\}\) can contribute to \(H_n\). The autocorrelation formula for \(C_m\) then gives \(C_m=0\) for \(|m|>2\).
\end{suppproof}

\begin{suppproof}{Proposition~\ref{prop:power_reconfigurability}: Power variation}
Termwise integration of the absolutely convergent series in~\eqref{eq:power_fourier_series} is valid. Since
\[
 \frac{1}{2\Delta}\int_{\psi_c-\Delta}^{\psi_c+\Delta}e^{jm\psi}d\psi
 =e^{jm\psi_c}\operatorname{sinc}(m\Delta),
\]
the stated expression for \(\bar P_\Psi\) follows directly.

The channel in~\eqref{eq:coupled_channel_swivel} is real analytic in \(\psi\), and hence so is \(P(\psi)\). If \(\sigma_{P,\Psi}^2=0\), continuity implies \(P(\psi)=\bar P_\Psi\) throughout the interval. Analytic continuation then makes \(P\) constant for all \(\psi\), which is equivalent to \(C_m=0\) for every \(m\neq0\). The converse is immediate from~\eqref{eq:power_fourier_series}.

For the lower bound in~\eqref{eq:selection_gain_bounds}, \(0\le P(\psi)\le P^\star\) gives
\[
 \sigma_{P,\Psi}^2+\bar P_\Psi^2
 =\frac{1}{2\Delta}\int_\Psi P^2(\psi)d\psi
 \le P^\star\bar P_\Psi.
\]
Dividing by \(\bar P_\Psi>0\) yields
\(
\sigma_{P,\Psi}^2/\bar P_\Psi\le P^\star-\bar P_\Psi
\).

For the upper bound, \(|g_\ell(\psi)|\le1\) and~\eqref{eq:srs_orientation_motion} give
\(
|g_\ell'(\psi)|\le\|d\bm n/d\psi\|=\sqrt{1-\mu}
\).
Also, the pairwise phase derivative satisfies
\(
|\vartheta_{\ell i}'(\psi)|\le r_{\ell i}
\).
Differentiating~\eqref{eq:power_decomposition} and applying these bounds gives
\[
 |P'(\psi)|
 \le2\sqrt{1-\mu}\left(\sum_{\ell=1}^{L}|\alpha_\ell|\right)^2
 +2\sum_{\ell<i}|\alpha_\ell\alpha_i|r_{\ell i}
 =B_P.
\]
Let \(\psi^\star\in\Psi\) attain \(P^\star\).
The Lipschitz bound implies \(P^\star-P(\psi)\le B_P|\psi^\star-\psi|\).
Averaging over \(\Psi\) and using
\(
(2\Delta)^{-1}\int_\Psi|\psi^\star-\psi|d\psi\le\Delta
\)
gives \(P^\star-\bar P_\Psi\le\Delta B_P\). For \(\Delta=\pi\), Fourier orthogonality gives \(\bar P_{2\pi}=C_0\) and Parseval's identity gives \(\sigma_{P,2\pi}^2=\sum_{m\neq0}|C_m|^2\).
\end{suppproof}

\begin{suppproof}{Theorem~\ref{thm:orientation_coverage}: Orientation guarantee}
For any nonzero complex field \(\bm f\), the selected power is at least its interval average:
\[
\begin{aligned}
 P^\star
 &\ge\frac{1}{2\Delta}\int_{-\Delta}^{\Delta}
 |\bm n(\tilde\psi)^{\mathsf T}\bm f|^2d\tilde\psi\\
 &=\bm f^{\mathsf H}\bm Q_\Delta(\mu)\bm f.
\end{aligned}
\]
The Rayleigh quotient therefore gives~\eqref{eq:orientation_coverage_bound}. The second coordinate is decoupled in \(\bm Q_\Delta(\mu)\), giving the eigenvalue \((1-\mu)V_\Delta\). The remaining two eigenvalues are those of the \((1,3)\) block, whose smaller eigenvalue is \(\lambda_-(\mu,\Delta)\) in~\eqref{eq:orientation_coverage_eigenvalue}. This proves the stated expression for \(\lambda_\Delta(\mu)\).

For \(0<\mu<1\) and \(\Delta>0\), suppose a real vector \(\bm x\) satisfies \(\bm x^{\mathsf T}\bm Q_\Delta\bm x=0\). Then \(\bm x^{\mathsf T}\bm n(\tilde\psi)=0\) for every \(\tilde\psi\) in an interval of positive length. Since this quantity is a linear combination of \(1\), \(\cos\tilde\psi\), and \(\sin\tilde\psi\), all three coefficients must vanish, so \(\bm x=\bm0\). Hence \(\bm Q_\Delta\) is positive definite and \(\lambda_\Delta(\mu)>0\).

For \(\Delta=\pi\), \(S_\pi=0\) and \(\operatorname{sinc}(2\pi)=0\), so
\[
 \bm Q_\pi=\operatorname{diag}\!\left(
 \frac{1-\mu}{2},\frac{1-\mu}{2},\mu\right).
\]
The lower bound is therefore \(\min\{\mu,(1-\mu)/2\}\). It is also attainable as a worst case. If \(\mu\le(1-\mu)/2\), choose \(\bm f\) parallel to \(\bm e\), for which the normalized power is constantly \(\mu\). If \((1-\mu)/2\le\mu\), choose a circularly polarized transverse field \(\bm f=(\bm e_1+j\bm e_2)/\sqrt2\), for which the normalized power is constantly \((1-\mu)/2\). This establishes~\eqref{eq:full_circle_complex_capability}.
\end{suppproof}

\begin{suppproof}{Proposition~\ref{prop:robust_orientation_mounting}: Optimal orientation guarantee}
For \(0<\Delta<\pi\), write \(U=U_\Delta\), \(S=S_\Delta\), and let \(q(\mu)=\lambda_-(\mu,\Delta)\). The eigenvalue \(q\) is the smaller root of
\[
 F(q,\mu)
 =\bigl((1-\mu)U-q\bigr)(\mu-q)
 -\mu(1-\mu)S^2=0.
\]
At an interior stationary point of the smaller eigenvalue branch, implicit differentiation gives \(F_\mu(q,\mu)=0\), where
\[
 F_\mu(q,\mu)
 =2S^2\mu-S^2+Uq-2U\mu+U-q.
\]
Solving these two equations and selecting the solution on the smaller eigenvalue branch gives
\[
 \mu^\star=\frac{U+S}{1+U+2S},
 \qquad
 q(\mu^\star)=\frac{U-S^2}{1+U+2S}.
\]
The other algebraic stationary solution belongs to the larger eigenvalue branch. Since \(q(0)=q(1)=0\) and \(q(\mu)>0\) for \(0<\mu<1\), this stationary point maximizes \(q(\mu)\).

It remains to verify that the decoupled eigenvalue \((1-\mu)V_\Delta\) does not become the active minimum at this point. Direct substitution gives
\[
 (1-\mu^\star)V_\Delta-q(\mu^\star)
 =\frac{V_\Delta(1+S_\Delta)+S_\Delta^2-U_\Delta}
 {1+U_\Delta+2S_\Delta}\ge0,
\]
for \(0<\Delta<\pi\). Hence \(\lambda_\Delta(\mu^\star)=q(\mu^\star)\), proving~\eqref{eq:complex_mu_opt} for this range. For \(\Delta=\pi\), the full-circle expression in the proof of Theorem~\ref{thm:orientation_coverage} gives
\(\lambda_\pi(\mu)=\min\{\mu,(1-\mu)/2\}\), which is maximized at \(\mu^\star=1/3\) with \(\lambda_\pi^\star=1/3\), consistent with~\eqref{eq:complex_mu_opt}.

For the small-interval behavior,
\[
\begin{aligned}
 S_\Delta&=1-\frac{\Delta^2}{6}+\frac{\Delta^4}{120}+O(\Delta^6),\\
 U_\Delta&=1-\frac{\Delta^2}{3}+\frac{\Delta^4}{15}+O(\Delta^6).
\end{aligned}
\]
Substitution into~\eqref{eq:complex_mu_opt} yields
\[
 \mu^\star(\Delta)=\frac12-\frac{\Delta^2}{24}+O(\Delta^4),
 \qquad
 \lambda_\Delta^\star=\frac{\Delta^4}{180}+O(\Delta^6),
\]
as stated.
\end{suppproof}

\begin{suppproof}{Corollary~\ref{cor:specified_linear_field}: Linear polarization matching}
The scalar projection can be written as
\[
\begin{aligned}
 \bm n(\tilde\psi)^{\mathsf T}\bm\epsilon
 &=a+b\cos(\tilde\psi-\varphi_f),\\
 a&=\sqrt{\mu\eta},\qquad
 b=\sqrt{(1-\mu)(1-\eta)}.
\end{aligned}
\]
Over \([-\Delta,\Delta]\), the largest attainable cosine is \(\cos d_0\), while the smallest is \(-\cos d_\pi\). Since the absolute value of an affine function over an interval is maximized at one of its two endpoint values, the selected squared projection is exactly the maximum stated in the corollary.

Perfect matching requires \(|\bm n^{\mathsf T}\bm\epsilon|=1\). Equality in Cauchy-Schwarz holds if and only if the two real unit vectors are parallel or antiparallel, giving the finite-interval condition. Over a full \(2\pi\) swivel range, the cosine can equal one, so
\[
 \frac{P^\star}{\|\bm f\|^2}
 =\left[\sqrt{\mu\eta}+\sqrt{(1-\mu)(1-\eta)}\right]^2\le1.
\]
The equality condition for the two-dimensional unit vectors \((\sqrt\mu,\sqrt{1-\mu})\) and \((\sqrt\eta,\sqrt{1-\eta})\) is \(\mu=\eta\).
\end{suppproof}

\begin{suppproof}{Proposition~\ref{prop:two_path_translation}: Two-path phase coverage}
For two paths, the power reduces to
\[
 P(\tilde\psi)=A_1^2+A_2^2
 +2A_1A_2\cos\bigl(\chi_{12}-\vartheta_{12}(\tilde\psi)\bigr).
\]
The continuous function \(\vartheta_{12}(\tilde\psi)\) attains exactly the interval \(\Theta_{12}\). Maximizing the cosine is therefore equivalent to choosing the attainable phase closest to \(\chi_{12}\) on the circle, which gives~\eqref{eq:two_path_exact_power}. Perfect coherent combining for the given phase offset occurs exactly when this circular distance is zero.

A contiguous real interval contains a representative of every phase modulo \(2\pi\) if and only if its width is at least \(2\pi\). Thus universal phase coverage is equivalent to \(W_{12}\ge2\pi\). Over a full swivel revolution, \(W_{12}=2k\rho\xi_{12}\). If \(\xi_{12}=0\), translation cannot change the relative phase. Otherwise,
\[
 2k\rho\xi_{12}\ge2\pi
 \quad\Longleftrightarrow\quad
 \rho\ge\frac{\lambda_c}{2\xi_{12}},
\]
which proves~\eqref{eq:full_circle_phase_radius}.
\end{suppproof}

\suppheading{Additional Analysis Near Full Extension}

The preceding analysis shows that wireless reconfigurability depends jointly on the antenna motion and the feasible swivel interval. Both are determined, in part, by the prescribed physical task. We now examine a limiting case that illustrates why the geometric size of the antenna trajectory alone is not a sufficient measure of usable wireless mobility.

Consider an SRS manipulator approaching full arm extension, where the shoulder-to-wrist distance satisfies $d\to (L_1+L_2)^-$. From~\eqref{eq:elbow_radius_task}, the elbow-circle radius approaches zero, i.e., $r_E\to0$. Consequently, $s_X\to0$, $c_X\to1$, and the centerline motion radius $\rho_{0,X}\to0$.

Nevertheless, an antenna mounted on the cylindrical link surface is offset from the link centerline by $\varrho_X$. This offset can preserve a nonzero radius of the antenna phase-center locus even as the elbow circle shrinks. The following result distinguishes this nominal geometric radius from the antenna motion that remains accessible under the task.

\begin{suppproposition}[Near full extension]
\label{prop:task_boundary}
Fix a surface mounting $\bm m=(X,\tau,\phi,\beta)$ with $\varrho_X>0$, and consider regular tasks with $d\to (L_1+L_2)^-$. Then $\rho(d)\to\varrho_X$ and $\mu(d)\to\cos^2\beta$, so the surface offset preserves a nonzero limiting phase center radius. If the feasible swivel half-width also satisfies $\Delta(d)\to0$, then
\begin{equation*}
\begin{aligned}
 \max_{\psi_1,\psi_2\in\Psi}
 \|\bm p_A(\psi_1)-\bm p_A(\psi_2)\|
 &=2\rho\Delta+O(\Delta^3),\\
 \max_{\psi_1,\psi_2\in\Psi}
 \|\bm n(\psi_1)-\bm n(\psi_2)\|
 &=2\sqrt{1-\mu}\,\Delta+O(\Delta^3),
\end{aligned}
\end{equation*}
Moreover, for any sequence of finite multipath channels with fixed $L$ and uniformly bounded path coefficients, $P^\star-\bar P_\Psi\to0$.
\end{suppproposition}

\begin{suppproof}{Proposition~\ref{prop:task_boundary}: Near full extension}
As \(d\to (L_1+L_2)^-\),~\eqref{eq:elbow_radius_task} gives \(r_E\to0\). Hence \(\rho_{0,X}\to0\), \(s_X\to0\), and \(c_X\to1\). Equations~\eqref{eq:surface_position_descriptors} and~\eqref{eq:surface_orientation_descriptors} then give
\(
\rho\to\varrho_X
\)
and
\(
\mu=\nu^2\to\cos^2\beta
\).

For sufficiently small \(\Delta\), the largest angular separation between two feasible swivel values is \(2\Delta\). The corresponding chord lengths of the phase center circle and the orientation cone are
\[
\begin{aligned}
 2\rho\sin\Delta
 &=2\rho\Delta+O(\Delta^3),\\
 2\sqrt{1-\mu}\sin\Delta
 &=2\sqrt{1-\mu}\,\Delta+O(\Delta^3).
\end{aligned}
\]
Finally, for fixed \(L\) and uniformly bounded path coefficients, \(B_P\) in~\eqref{eq:power_slope_bound} remains uniformly bounded because \(\rho\) has a finite limit and the propagation directions are unit norm. The upper bound in~\eqref{eq:selection_gain_bounds} therefore gives \(P^\star-\bar P_\Psi\le\Delta B_P\to0\).
\end{suppproof}

Proposition~\ref{prop:task_boundary} reveals an important distinction between nominal antenna mobility and usable wireless reconfiguration. Even when the phase center retains a nonzero circular motion radius, the actual position and orientation excursions become small if the feasible swivel interval collapses. The resulting received-power improvement above the interval mean also vanishes.

The conclusion that the gain vanishes requires the feasible swivel span to shrink. Approaching full extension alone does not imply that $\Delta$ must approach zero. Rather, the result highlights the role of task-dependent mechanical feasibility: the radius of the possible antenna trajectory is not enough to determine how much of that trajectory communication can exploit.

\suppheading{Additional Mounting and Approximation Analysis}

\suppsubheading{Attainable Motion Descriptors}

The mounting $\bm m=(X,\tau,\phi,\beta)$ determines how the host's self-motion is converted into antenna translation and orientation. Recall from Section~\ref{sec:reconfigurability} that two useful geometric descriptors are the phase-center circle radius $\rho$ and the orientation-cone parameter $\mu=\nu^2$. The former determines the magnitude of antenna displacement per unit swivel, while the latter determines the axial and transverse components of the dipole orientation.

To characterize the values of $(\rho,\mu)$ attainable through the mounting geometry, consider a fixed prescribed task, supporting link $X\in\{U,F\}$, and circumferential angle $\phi$. From~\eqref{eq:surface_position_descriptors},
\begin{equation*}
  \rho^2=
  \bigl(\rho_{0,X}+\varrho_Xc_X\cos\phi\bigr)^2
  +\varrho_X^2\sin^2\phi.
\end{equation*}
As the axial mounting coordinate $\tau$ varies from $0$ to $1$, the centerline radius $\rho_{0,X}$ covers $[0,r_E]$. Thus, the attainable phase-center radius is determined by a convex quadratic function of $\rho_{0,X}$.

Similarly, the orientation parameter can be controlled through the dipole mounting angle $\beta$. When $1-s_X^2\cos^2\phi>0$, define
$\beta_{0,X}(\phi)=\operatorname{atan2}(\sigma_Xs_X\sin\phi,c_X)$.
Equation~\eqref{eq:surface_orientation_descriptors} can then be written as
$\nu=\sqrt{1-s_X^2\cos^2\phi}\cos[\beta-\beta_{0,X}(\phi)]$.
When $1-s_X^2\cos^2\phi=0$, we have $\nu=0$ for every $\beta$. These expressions give the following characterization.

\begin{suppproposition}[Attainable motion descriptors]
\label{prop:reachable_motion_descriptors}
Fix a prescribed task, a supporting link $X\in\{U,F\}$, and a circumferential mounting angle $\phi$. When $\tau\in[0,1]$ and $\beta\in[0,\pi]$, the attainable pairs $(\rho,\mu)$ form exactly the set
\begin{equation*}
  (\rho,\mu)\in
  [\rho_{\min,X}(\phi),\rho_{\max,X}(\phi)]
  \times[0,\mu_{\max,X}(\phi)],
\end{equation*}
where
\begin{equation*}
\begin{aligned}
  \rho_{\min,X}^2(\phi)
  &=
  \varrho_X^2\sin^2\phi
  +\operatorname{dist}^2
  \left(
  -\varrho_Xc_X\cos\phi,[0,r_E]
  \right),\\
  \rho_{\max,X}^2(\phi)
  &=
  \varrho_X^2\sin^2\phi
  +\max_{x\in\{0,r_E\}}
  (x+\varrho_Xc_X\cos\phi)^2,\\
  \mu_{\max,X}(\phi)
  &=1-s_X^2\cos^2\phi.
\end{aligned}
\end{equation*}
Here, $\operatorname{dist}(x,\mathcal S)$ denotes the Euclidean distance from $x$ to the set $\mathcal S$.
\end{suppproposition}

\begin{suppproof}{Proposition~\ref{prop:reachable_motion_descriptors}: Attainable motion descriptors}
For either link, \(\rho_{0,X}\) spans \([0,r_E]\) as \(\tau\) spans \([0,1]\). Setting \(x=\rho_{0,X}\),~\eqref{eq:surface_position_descriptors} gives
\[
 \rho^2(x)=\bigl(x+\varrho_Xc_X\cos\phi\bigr)^2
 +\varrho_X^2\sin^2\phi,
 \qquad 0\le x\le r_E.
\]
This is a convex quadratic. Its minimum is obtained by projecting the unconstrained minimizer \(-\varrho_Xc_X\cos\phi\) onto \([0,r_E]\), which gives the distance expression for \(\rho_{\min,X}^2\). A convex function on a compact interval attains its maximum at an endpoint, giving \(\rho_{\max,X}^2\).

For orientation, when \(\mu_{\max,X}(\phi)>0\), the relation preceding the proposition gives
\(\mu=\mu_{\max,X}(\phi)\cos^2[\beta-\beta_{0,X}(\phi)]\).
As \(\beta\) spans an interval of length \(\pi\), the squared cosine spans \([0,1]\), so \(\mu\) spans \([0,\mu_{\max,X}(\phi)]\). If \(\mu_{\max,X}(\phi)=0\), then \(\mu=0\) for all \(\beta\). Finally, \(\tau\) and \(\beta\) are independent mounting variables for fixed \((X,\phi)\), so every pair in the Cartesian product of the two ranges is attainable.
\end{suppproof}

Proposition~\ref{prop:reachable_motion_descriptors} describes the range of antenna motion geometries that can be obtained by changing the physical mounting. For fixed $(X,\phi)$, the axial coordinate $\tau$ controls the attainable circle radius, while the dipole angle $\beta$ controls the orientation-cone parameter. Mechanical restrictions on admissible mounting locations can further reduce these ranges.

The characterization also connects the mounting variables to the capability results in Section~\ref{sec:reconfigurability}. For example, the orientation-only guarantee in Proposition~\ref{prop:robust_orientation_mounting} is maximized at $\mu^\star(\Delta)$. At a given $(X,\phi)$ with $\mu_{\max,X}(\phi)>0$, this target can be realized if
$\mu^\star(\Delta)\le\mu_{\max,X}(\phi)$ and an admissible $\beta$ satisfies
$\cos^2[\beta-\beta_{0,X}(\phi)]=\mu^\star(\Delta)/\mu_{\max,X}(\phi)$.
If $\mu_{\max,X}(\phi)=0$, only $\mu=0$ is attainable. For translation, the capability depends on the circle radius $\rho$, but also on the alignment of the feasible arc through $\delta_p$, as established in Proposition~\ref{prop:two_path_translation}.

These separate design insights do not, however, determine the best mounting for the complete wireless channel. Although $(\rho,\mu)$ characterize the magnitudes of position and orientation variations, the coupled response also depends on the circle center $\bm p_c$, the signed axial component $\nu$, and the angular offsets $\delta_p$ and $\delta_n$. Moreover, these quantities depend on the prescribed task through the robot geometry.

\begin{remark}[Geometric mobility versus wireless utility]
\label{rem:mounting_capability}
The ranges in Proposition~\ref{prop:reachable_motion_descriptors} describe alternative physical mountings, not independent online control of $\rho$ and $\mu$. Once a mounting is installed, its motion descriptors are fixed for a given task, and communication can adjust only $\psi\in\Psi$. Furthermore, a larger circle radius or a more favorable orientation cone does not necessarily yield higher received power, because the two mechanisms are coupled and their wireless effects depend on the propagation environment. Mounting design must therefore evaluate the complete channel response rather than optimize $\rho$ and $\mu$ separately.
\end{remark}

\suppsubheading{Existence and SAA Consistency}

The cross-task mounting objective admits a maximizer under the following finite channel-moment condition. Suppose that, for every task with $\pi_a>0$,
\begin{equation*}
  \mathbb E_{\mathcal H_a}
  \left[
    \left(
      \sum_{\ell=1}^{L}
      |\bar\alpha_{a,\ell}|
    \right)^2
  \right]<\infty.
\end{equation*}
For each channel realization, $P_a(\psi;\bm m,\mathcal H_a)$ is continuous in the mounting variables. Since $\Psi_a$ is compact and independent of $\bm m$, the selected power $P_a^\star(\bm m,\mathcal H_a)$ is also continuous in $\bm m$. Moreover,
\begin{equation*}
  P_a^\star(\bm m,\mathcal H_a)
  \le
  \left(
    \sum_{\ell=1}^{L}
    |\bar\alpha_{a,\ell}|
  \right)^2,
\end{equation*}
which provides an integrable upper bound. Consequently, $J(\bm m)$ is continuous on the compact set $\mathcal M$, and a solution to~\eqref{eq:joint_mounting_problem} exists.

The corresponding SAA mounting problem is
$\widehat{\bm m}_{N_c}\in\arg\max_{\bm m\in\mathcal M}\widehat J_{N_c}(\bm m)$.
Under the finite-moment condition and independent sampling for each task, standard SAA consistency results on the compact set $\mathcal M$ apply~\cite{shapiro_stochastic_2014}:
\[
  \sup_{\bm m\in\mathcal M}
  \left|
    \widehat J_{N_c}(\bm m)-J(\bm m)
  \right|
  \xrightarrow[N_c\to\infty]{\mathrm{a.s.}}0.
\]
Hence, every accumulation point of SAA maximizers belongs almost surely to
$\arg\max_{\bm m\in\mathcal M}J(\bm m)$.
If the optimum is unique, then
$\widehat{\bm m}_{N_c}\to\bm m^\star$
almost surely. Thus, the sample-based design approaches the statistical design as the number of channel samples increases.

\suppsubheading{Uniform Swivel-Grid Approximation}

Since the grid is a subset of the feasible interval, the grid-selected power cannot exceed the continuous selected power. The remaining question is how much performance can be lost by restricting the selection to grid points. The power slope bound in~\eqref{eq:power_slope_bound} provides an explicit answer.

For any fixed task $a$, mounting $\bm m$, and channel
$\mathcal H_a$, the power slope bound $B_{P,a}$
in~\eqref{eq:power_slope_bound} gives
\begin{equation}
\begin{aligned}
 0&\le P_a^\star(\bm m,\mathcal H_a)
 -\widetilde P_{a,Q}^\star(\bm m,\mathcal H_a)\\
 &\le\frac{\delta_{\psi,a}}{2}B_{P,a}(\bm m,\mathcal H_a).
\end{aligned}
\label{eq:swivel_grid_error}
\end{equation}
The bound decreases with the swivel-grid spacing. Its uniform extension
also bounds the mounting optimization loss from optimizing the grid-based SAA
objective, as established in supplementary Proposition~\ref{prop:swivel_grid_approximation}.

The following result extends the pointwise bound uniformly over feasible physical mountings and quantifies its effect on mounting optimization.

\begin{suppproposition}[Uniform swivel-grid approximation]
\label{prop:swivel_grid_approximation}
Let $\rho_{\mathcal M}=\max_{1\le a\le N_T}\max_{\bm m\in\mathcal M}\rho_a(\bm m)$.
The following bound holds uniformly over $\mathcal M$:
\begin{equation*}
  0\le
  \widehat J_{N_c}(\bm m)
  -\widetilde J_{N_c,Q}(\bm m)
  \le\varepsilon_Q,
\end{equation*}
where
\begin{equation*}
\begin{aligned}
  \varepsilon_Q
  ={}&
  \sum_{a=1}^{N_T}
  \frac{\pi_a\delta_{\psi,a}}{2N_c}
  \sum_{r=1}^{N_c}
  \Bigg[
    2\left(
      \sum_{\ell=1}^{L}
      |\bar\alpha_{a,r,\ell}|
    \right)^2\\
  &\qquad\qquad
  +4k\rho_{\mathcal M}
    \sum_{1\le\ell<i\le L}
    |\bar\alpha_{a,r,\ell}
      \bar\alpha_{a,r,i}|
  \Bigg].
\end{aligned}
\end{equation*}

Here, $\bar\alpha_{a,r,\ell}$ denotes the coefficient of path $\ell$ in the channel sample $\mathcal H_{a,r}$.

If $\widetilde{\bm m}_{N_c,Q}$ globally maximizes~\eqref{eq:grid_saa_objective}, then
\begin{equation*}
\begin{aligned}
  &\widehat J_{N_c}
  (\widehat{\bm m}_{N_c})
  -\widehat J_{N_c}
  (\widetilde{\bm m}_{N_c,Q})\\
  &\qquad\le\varepsilon_Q.
\end{aligned}
\end{equation*}
\end{suppproposition}

\begin{suppproof}{Proposition~\ref{prop:swivel_grid_approximation}: Grid approximation}
For the uniform SAA bound, the center-referenced coefficient used in Section~\ref{sec:response} differs from \(\bar\alpha_{a,r,\ell}\) only by a unit-modulus phase factor. Hence their magnitudes are equal. Moreover,
\(
\sqrt{1-\mu_a(\bm m)}\le1
\),
\(
\rho_a(\bm m)\le\rho_{\mathcal M}
\), and the projected difference of two unit propagation directions has norm at most two. Therefore
\[
 B_{P,a}
 \le2\left(\sum_{\ell=1}^{L}|\bar\alpha_{a,r,\ell}|\right)^2
 +4k\rho_{\mathcal M}
 \sum_{\ell<i}|\bar\alpha_{a,r,\ell}\bar\alpha_{a,r,i}|.
\]
Multiplying the single-sample error by \(\pi_a/N_c\) and summing over tasks and channel samples yields
\(0\le\widehat J_{N_c}(\bm m)-\widetilde J_{N_c,Q}(\bm m)\le\varepsilon_Q\),
uniformly in \(\bm m\).

Let $\widehat{\bm m}_{N_c}$ and $\widetilde{\bm m}_{N_c,Q}$
be global maximizers of the continuous and grid-based SAA objectives,
respectively. The grid is a subset of the feasible interval, hence
$\widetilde J_{N_c,Q}(\bm m)\le\widehat J_{N_c}(\bm m)$
for every mounting. Combining this one-sided bound with the optimality
of $\widetilde{\bm m}_{N_c,Q}$ gives
\[
\begin{aligned}
 &\widehat J_{N_c}(\widehat{\bm m}_{N_c})
 -\widehat J_{N_c}(\widetilde{\bm m}_{N_c,Q})\\
 &\le\widehat J_{N_c}(\widehat{\bm m}_{N_c})
 -\widetilde J_{N_c,Q}(\widetilde{\bm m}_{N_c,Q})\\
 &\le\widehat J_{N_c}(\widehat{\bm m}_{N_c})
 -\widetilde J_{N_c,Q}(\widehat{\bm m}_{N_c})\\
 &\le\varepsilon_Q.
\end{aligned}
\]
This proves the bound on the mounting optimization loss.
\end{suppproof}

\end{document}